\documentclass{aastex}
\usepackage{natbib,epsfig,emulateapj5,rotate}

\tightenlines

\newcommand{\be}{\begin{eqnarray}}
\newcommand{\ee}{\end{eqnarray}}

\newcommand{\ssys}{S_{\rm sys}}

\newcommand{\npol}{N_{\rm pol}}

\newcommand{\DM}{{\rm DM}}
\newcommand{\DtDM}{\Delta t_{\DM}}
\newcommand{\DtdDM}{\Delta t_{\delta \DM}}
\newcommand{\DtDnu}{\Delta t_{\rm \Delta\nu}}
\newcommand{\DtISS}{\tau_d}
\newcommand{\Dnu}{\Delta\nu}
\newcommand{\dDM}{\delta\DM}

\newcommand{\Dnuiss}{\Delta\nu_{\rm d}}
\newcommand{\Dtiss}{\Delta t_{\rm d}}
\newcommand{\SM}{{\rm SM}}
\newcommand{\thetad}{\theta_{d}}
\newcommand{\taud}{\tau_{d}}
\newcommand{\thetaiso}{\theta_{\rm iso}}
\newcommand{\ld}{\ell_{d}}
\newcommand{\thetaisor}{\theta_{\rm iso}}
\newcommand{\lr}{\ell_{r}}
\newcommand{\dkpc}{D_{\rm kpc}}
\newcommand{\viss}{V_{\rm ISS}}

\newcommand{\vsperp}{{V_{\rm s}}_{\perp}}
\newcommand{\Niss}{N_{\rm ISS}}

\begin{document}
\title{Searches for Fast Radio Transients}
\author{J.\ M.\ Cordes\altaffilmark{1} \& M.\ A.\ McLaughlin\altaffilmark{2}}
\altaffiltext{1}{Astronomy Department and NAIC, Cornell University, Ithaca, NY 14853}
\altaffiltext{2}{Jodrell Bank Observatory, University of Manchester, Macclesfield, Cheshire, SK11 9DL, UK}

\begin{abstract}
We discuss optimal detection of fast radio transients from
astrophysical objects
while taking into account the effects of propagation through
intervening ionized media, including
dispersion, scattering and scintillation.
Our analysis applies to the giant-pulse phenomenon
exhibited by some  pulsars, for which we show examples,
and to radio pulses from other astrophysical
sources, such as prompt radio emission from gamma-ray burst sources and
modulated signals from extra-terrestrial civilizations. 
We estimate scintillation parameters for extragalactic sources 
that take into account scattering in both the host galaxy and
in foreground, Galactic plasma.
\end{abstract}
\keywords{radio pulses, pulsars, interstellar medium, dispersion, scattering, gamma-ray
 bursts, SETI}

\section{Introduction}

Fast radio transients -- pulses with durations of seconds or less -- are
important indicators of coherent emission processes in astrophysical
sources. The detection of such pulses is complicated by their propagation
through the intervening ionized media.   We focus on 
the interstellar medium (ISM) in this paper, but the intergalactic
and interplanetary media are also relevant for some objects.
The dominant effects,
dispersion and scattering, distort pulses but also provide a
signature for distinguishing them from terrestrial radio-frequency
interference (RFI). In this paper we discuss algorithms for detecting
pulses that are dispersed and scattered to varying degrees by the ISM.
This discussion is quite general. Though it is motivated by pulsar radio
emission, particularly the so-called `giant-pulses' from the Crab pulsar
and a few other objects, our analysis applies to any other transient radio
source.

Searches for radio transients have a long history. Indeed, pulsars were
first discovered by the detection of single pulses on a chart recorder
\cite{hewish68}, with the Crab pulsar detected through its giant pulses
soon after \cite{sta68}. Davies \& Large (1970) detected three new pulsars
in their single-pulse search but failed to detect any pulses from 13
supernova remnants. A search for single pulses from the X-ray sources
Scorpius X-1 and Cygnus X-1, at one time thought to be possibly associated
with pulsars, was also unsuccessful \cite{taylor72}. In a search for radio
counterparts to the gravitational pulses reported by Weber (1969), both
Hughes \& Retallack (1973) and Edwards et al. (1974) detected excesses of radio
pulses from the direction of the Galactic center, but did not believe them
to be correlated with the gravitational pulses. With the goal of detecting
the single, large radio pulse expected to be emitted at the time of a
supernova explosion \cite{colgate71}, Huguenin \& Moore (1974) and
Kardashev et al. (1977) performed radio searches for single pulses, but
found no convincing signals of extraterrestrial origin aside from solar
flares. O'Sullivan et al. (1978) and Phinney \& Taylor (1979) conducted
searches for radio bursts possibly associated with annihilating black
holes, as suggested by Rees (1977), but likewise found no convincing
signals. Linscott \& Erkes (1980) reported the detection of highly
dispersed pulses from the elliptical galaxy M87, at a distance of roughly
15 Mpc, attributing these pulses to a massive compact object at the
center. However, subsequent attempts by others to confirm their findings
were unsuccessful \cite{hankins81,mcculloch81,taylor81}. In searches for
radio pulses associated with gamma-ray bursts, Cortiglioni et al. (1981),
Inzani et al. (1982) and Amy et al. (1989) detected some dispersed radio
pulses, but found no convincing associations with gamma-ray burst sources.
Vaughan \& Large (1989)  also conducted an unsuccessful search for radio
pulses from the soft gamma-ray repeater 0526$-$66. More recently, Nice (1999)
searched for radio pulses along 68 deg$^{2}$ of the Galactic plane at
430~MHz. This search resulted in the detection of individual pulses from 5
known pulsars and the discovery of one new pulsar which was previously
missed in a standard periodicity search \cite{nice95}. This pulsar
(J1918+08) does not emit giant pulses\footnote{We define giant pulses
as those comprising a long tail on the overall pulse amplitude distribution;
for the Crab pulsar and the millisecond pulsars B1937+21 and B1821-24,
these distributions are power-law in form.}
but is a normal, slow pulsar which
fortuitously emitted one strong pulse during the search observations.

The modest success of these searches for transient radio signals has been
largely due to limited sensitivity and sky coverage and, most importantly,
to the difficulty in discriminating between signals of astrophysical
origin and RFI. Array instruments such as the Low Frequency Array (LOFAR)
and the Square Kilometer Array (SKA) will be a vast improvement over
single-dish systems in all of these areas. With these arrays in the
developmental stages, it is timely to revisit searches for single pulses.
The outline of this paper is as follows. In \S\ref{sec:sources}, we
discuss the objects that might be sources of transient radio signals. In
\S\ref{sec:methods}, we describe single-pulse search algorithms and
present example results for giant-pulse emitting pulsars. We discuss
dispersion and pulse broadening from  scattering,  their effects on 
searches for single pulses, and ways to optimize searches in the 
presence of these effects. 
In \S\ref{sec:iss}, the effects of scintillations on source detection
are outlined.
As the increasing prominence of RFI makes it difficult to identify
transient pulses,   we briefly discuss RFI issues 
in \S\ref{sec:RFI}. 
Conclusions and a look to the future, especially for
RFI mitigation, are offered in
\S\ref{sec:conclusions}. In Paper II (McLaughlin \& Cordes 2003), 
we explicitly discuss giant pulses from Galactic pulsars and our 
attempts to detect similar pulses from nearby galaxies. 
In that paper we also outline  the conditions required for single-pulse
searches to be more effective than periodicity searches for
pulsars.

\section{Sources of Fast Radio Transients} \label{sec:sources}

While all pulsars show pulse-to-pulse intensity variations \cite{hesse74},
some pulsars have been found to emit so-called `giant' pulses, pulses with
strengths 100 or even 1000 times the mean pulse intensity. The Crab
(PSR~B0531+21) was the first pulsar found to exhibit this phenomenon. In
one hour of observation, the largest measured peak pulse flux of the Crab
is roughly $\sim$~10$^{5}$~Jy at 430~MHz for a duration of roughly
100~$\mu$s \cite{hankins75}, corresponding to an implied brightness 
temperature of 10$^{31}$ K. Recently, pulses with flux $\sim$ $10^{3}$~Jy 
at 5~GHz for a duration of only 2~ns have been detected from the 
Crab \cite{hankins03}. These `nano-giant' pulses imply brightness 
temperatures of 10$^{38}$~K, by far the most luminous emission from any 
astronomical object. For many years, this phenomenon was thought to be  
uniquely characteristic of the Crab. However, giant pulses have since been 
detected from the millisecond pulsars PSR~B1937+21 and PSR~B1821$-$24. From 
both of these pulsars, the largest measured peak pulse flux in one hour is
$\sim$ 10$^{3}$~Jy at 430~MHz \cite{cognard96,romani2001}.  Recently,
giant pulses have been detected from the Crab-like pulsar in
the Large Magellanic Cloud, PSR~B0540-69 (Johnston \& Romani, 2003).

Although pulsars may be the most well-studied transient radio sources,
they are by no means the only ones. In Figure~\ref{fig:temp_all}, we
compare the brightness temperatures of giant pulses with those from other
astrophysical sources of radio pulses. These sources include
brown dwarfs such as BD~LP944$-$20, from which radio
flares have recently been detected \cite{berger01}, and Jupiter, which has
long been known to emit radio flares at decameter wavelengths
\cite{aubier2000,lec1998}. Type II and Type III solar bursts are regularly
detected at radio frequencies of tens of MHz \cite{mann1996,poq1988}.
Radio flares from active stars such as UV Ceti and AD Leo are observed at
frequencies $\sim 1$ GHz \cite{jackson1989}. The emission from OH masers can
vary on timescales of hundreds of seconds and be detected as long-duration
radio bursts \cite{cohen1985,yud1986}. Active galactic nuclei (AGN)
outbursts, likely due to propagation of shocks in relativistic jets, are
observed at millimeter and centimeter wavelengths \cite{aller85,lain1994}.
Intraday variability (IDV) of other extragalactic sources, most likely
caused by interstellar scintillation, could also be detectable in searches
for radio transients \cite{ked2001}. 
Gamma-ray burst (GRB) afterglows are also modulated
by refractive, and possibly diffractive, scintillation \cite{goodman1997}.
Radio bursts from supernovae, as proposed by Colgate (1971) and detected
from supernova 1987a \cite{turtle1987}, bursts from the explosions of
primordial black holes \cite{phinney79}, and emission from
inspiraling neutron stars (Hansen \& Lyutikov 2000) 
could also be sources of fast
radio transients. Finally, searches for radio transients could detect
scintillation-modulated signals from extraterrestrial intelligent
civilizations \cite{cordes1997}.

\medskip
\epsfxsize=9truecm
\epsfbox{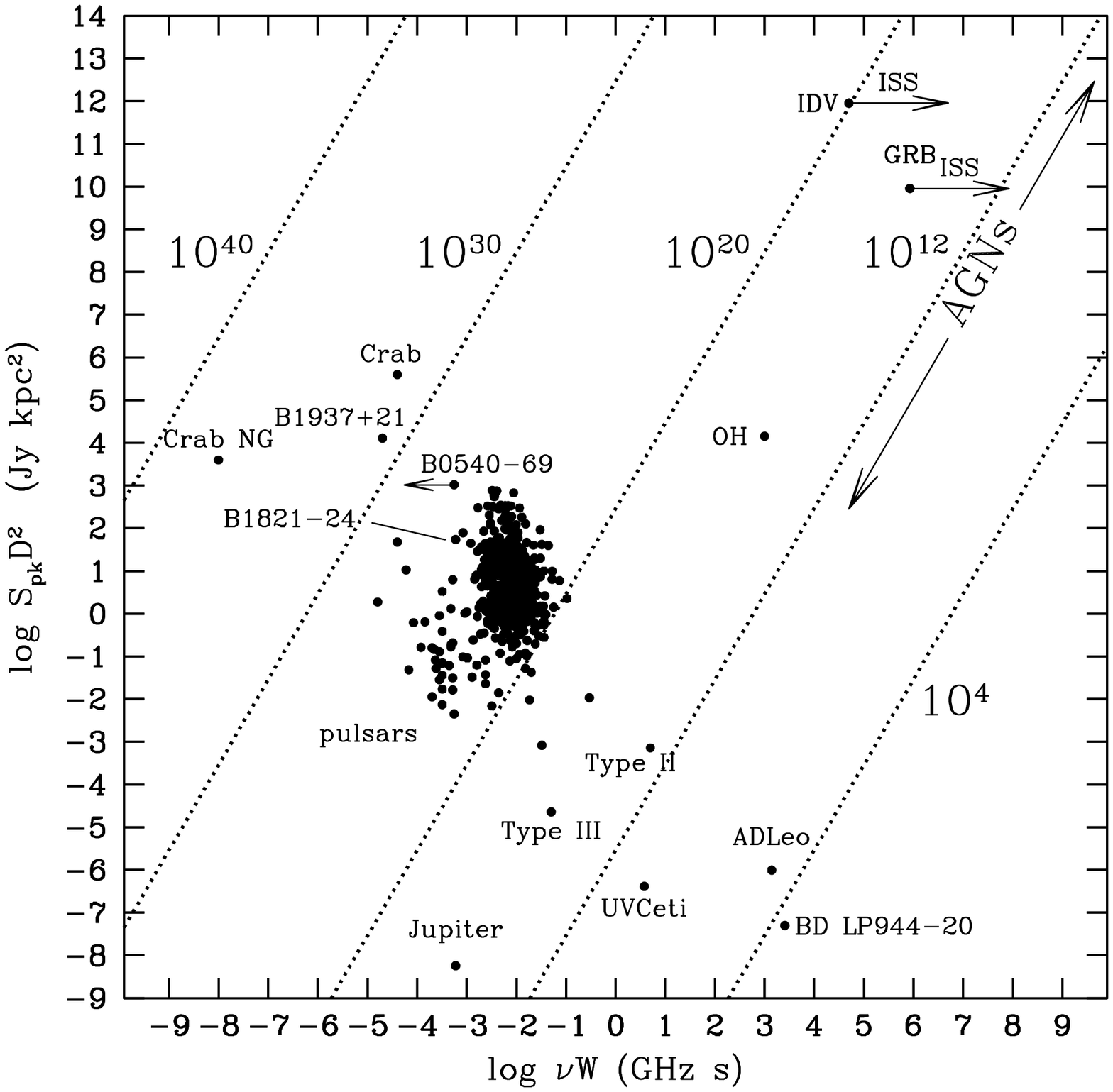}
\figcaption{\label{fig:temp_all}
A log-log plot of the product of peak flux $S$ in Jy and the square of the
distance $D$ in kpc vs. the product of frequency $\nu$ in GHz and pulse
width $W$ in s.
Points are shown for the `nano-giant' pulses detected from the Crab
(Hankins et al. 2003), the
giant pulses detected from the Crab, PSR~B1937+21 and
PSR~B1821$-$24, single pulses from all pulsars with flux, distance and
pulse width listed in the Princeton Pulsar Catalog \cite{ppcat} and other
possible sources of radio bursts.  Lines of constant brightness
temperature $T = S D^{2}/2k(\nu W)^{2}$ are shown, where $k$ is
Boltzmann's constant.  See text for explanations of particular plotted
symbols.
}
\bigskip

\section{Single-Pulse Search Methodology} \label{sec:methods}

In this section, we describe general issues that must be considered in
designing and carrying out any search for transient radio signals. We
first explore the various factors influencing the intensities,
 shapes and widths of
detected radio pulses. We then outline the steps of a search for dispersed
radio pulses, including dedispersion, matched filtering, thresholding and
diagnostics to determine the reality of detected signals. Lastly, we
describe the effects that interstellar scattering and scintillation have
on searches for fast radio transients and how to optimize searches in the
presence of these effects.

\subsection{Signal Model}

An appropriate model\footnote{ This model is only approximate.   
Receiver noise is strictly
additive only in the limit of large signal-to-noise ratio
and large time-bandwidth product.
Also,  the diffractive modulation $g_{\rm d}$  can be factored
as shown only if it is constant over the measurement bandwidth.
One may consider channelized signals for which this is true,
in which case the total intensity will be the sum over
channels of intensities like those given in the model. 
We also point out that when the diffractive scintillation
modulation $g_{\rm d}$ is important (i.e. not quenched by
bandwidth smoothing) multipath-induced pulse smearing is likely to be negligible
and vice versa.}    
for the measured intensity from a broadband source is
\be
I(t) &=& g_{\rm r} g_{\rm d} S(t) 
	* h_{\rm DM}(t) * h_{\rm d}(t) * h_{\rm Rx}(t) + N(t), 
\label{eq:sigmod}
\ee
where the source contribution, $S(t)$, is modulated by 
factors $g_{\rm r}$ and $g_{\rm d}$ that correspond to 
refractive and diffractive scintillation modulations (see below)
while also
being convolved (denoted by the asterisks) 
with several factors that account for
dispersion smearing, $h_{\rm DM}(t)$, pulse-broadening from
multipath propagation, $h_{\rm d}(t)$, and averaging
in the receiver and data acquisition system, $h_{\rm Rx}(t)$.
Additive receiver noise is represented by $N(t)$.
To a large
extent, dispersion smearing associated with
$h_{\rm DM}(t)$ can be deconvolved from the signal. 
However, there can be residual dedispersion smearing if
the exact value of dispersion measure (\DM) is not used,
as we discuss below.   
For large path lengths through the Galaxy and observations
at low frequencies,  pulse broadening from scattering  
can dominate the shape of
the measured pulse, in which case optimal detection entails usage of
an asymmetric matched filter given by the appropriate
pulse broadening function, $h_{\rm d}(t)$.
Below, we discuss the detection of broadband pulses with respect 
to these various factors. 

For narrowband signals, such as those that are hypothesized from 
extraterrestrial civilizations, the signal model in
Eq.~\ref{eq:sigmod} is applicable, though $S(t)$ now must be viewed
as a time modulation that also depends on radio frequency.   The
interplay of propagation effects with the intrinsic
properties of the signal need to be considered for particular cases.
In this paper we address broadband pulses explicitly
and simply state that many of our conclusions will also apply
to modulated, narrowband signals.   The effects of scintillations
in particular  are discussed in Cordes \& Lazio (1991) and
Cordes, Lazio \& Sagan (1997).

\subsection{Effective Time Resolution} \label{sec:timeres}

The shape of a detected radio pulse is influenced by a variety of factors.
A delta-function pulse emitted at the source will be detected with 
finite width due to a combination of propagation effects and
signal-processing response times. The effective time resolution for a
pulse is approximately the quadratic sum of the dispersion smearing, the
dedispersion error, the receiver filter response time and scattering
smearing, so that
\be
\Delta t =
        \left [
                \DtDM^2 + \DtdDM^2 +  \DtDnu^2  + \DtISS^2
        \right ]^{1/2}.
\label{eq:timeres}
\ee
The first term, the frequency-dependent smearing due to dispersion, is
equal to
 \be
\DtDM = 8.3 \,{\rm \mu \, s} \, {\rm DM} \, \Dnu_{\rm MHz}
     \nu_{\rm GHz}^{-3}
\label{eq:DtDM}
\ee
for a bandwidth $\Dnu_{\rm MHz}$, an observing frequency $\nu_{\rm GHz}$
and a dispersion measure DM, where DM is defined as the column
density along the line of sight (LOS) to a source (i.e. DM =
$\int_{0}^{\rm D} n_{e}(l) {\rm d}l$, where $n_e$ is the electron density
and D is the distance to the source) and has units
of pc cm$^{-3}$. Optimal detection requires that dispersion smearing be
removed from the measured signal by compensating time delays or,
equivalently, unwrapping phases of the Fourier components.  For any such
`dedispersion' technique, using a value for DM that is in error by an
amount $\dDM$ yields smearing of
\be
\DtdDM =
	\DtDM (\dDM/\DM).
\label{eq:DtdDM}
\ee
The third term, the filter response of an individual frequency channel, is
\begin{equation}
\DtDnu \sim (\Delta\nu)^{-1} =
        (\Delta\nu_{\rm MHz})^{-1}
        {\rm \mu \, s}.
\label{eq:DtDnu}
\end{equation}
Finally, the fourth term, $\DtISS$, is the broadening due to multipath
scattering, which has the effect of convolving the intrinsic pulse profile
with a function approximately exponential in form and having a
characteristic time scale $\tau_d$. Though it is highly dependent on
direction and distance, it may be estimated approximately as a statistical
function of DM.  The mean of this dependence is given by the empirical fit
(Cordes \& Lazio 2002)
\be
\log \taud
   = -3.72 + 0.411\log {\rm DM} + 0.937 (\log {\rm DM})^2 \nonumber \\
   - 4.4\log\nu_{\rm GHz} \, \mu s,
\label{eq:taud}
\ee
with scatter about the mean fit of $\sigma_{\log\tau_d} \approx 0.65$. The
coefficient 4.4 in the last term is based on the assumption of a
Kolmogorov spectrum for the scattering irregularities. For some lines of
sight the scaling with frequency may be weaker (e.g.  L{\" o}hmer et al.
2001), implying a coefficient closer to 4.0.  In Figure~\ref{fig:plottau3}
we illustrate some typical ISS parameters for pulsars. We further discuss
pulse broadening along with interstellar scintillation modulations in
\S\ref{sec:iss}, \S\ref{sec:expected} and in the Appendix.

\medskip
\epsfxsize=9truecm
\epsfbox{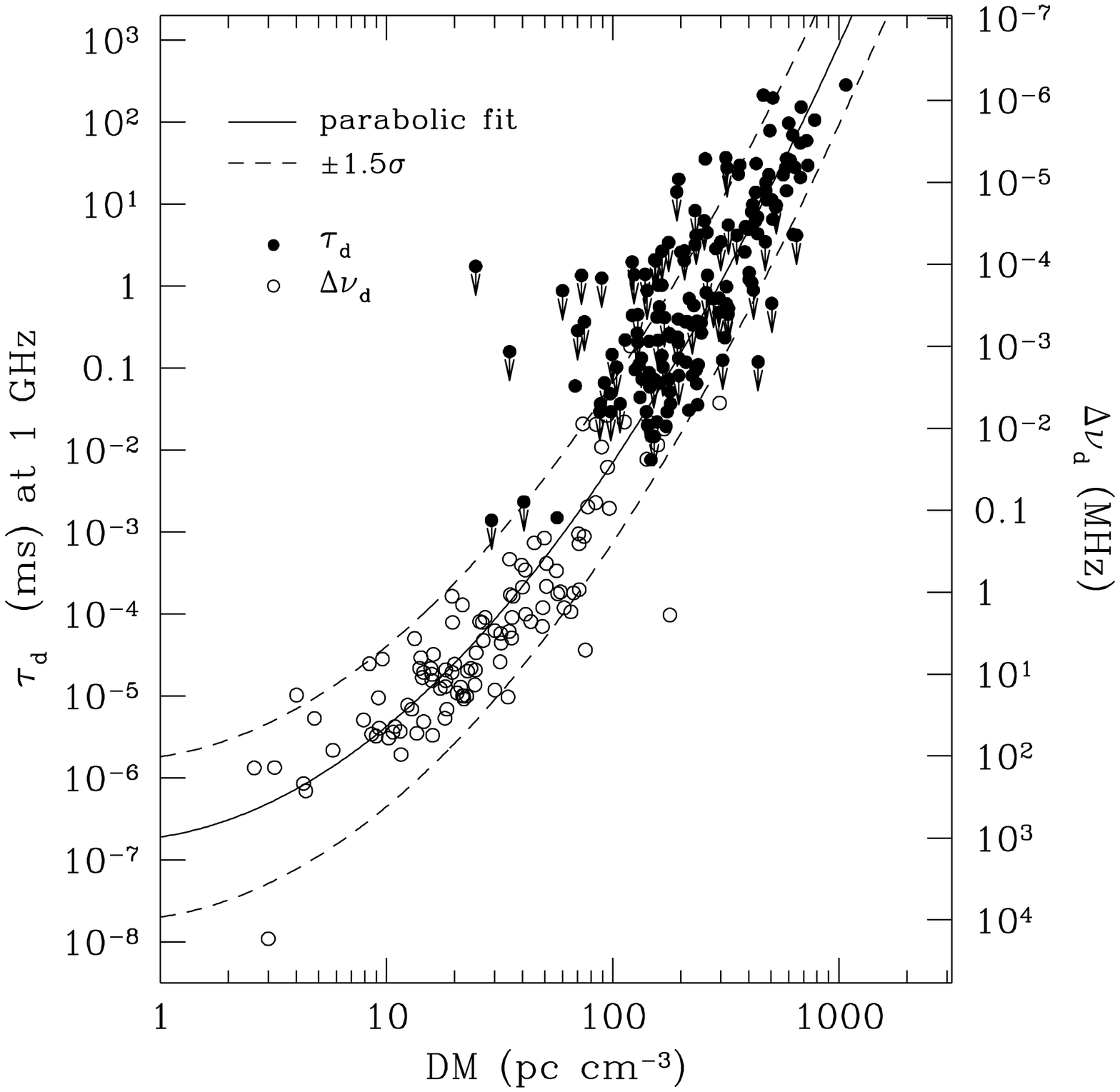}
\figcaption
{
Pulse broadening time $\tau_d$ and scintillation bandwidth $\Dnuiss$
plotted against DM (see Cordes \& Lazio 2002 for references).  The open
circles represent measurements of $\tau_d$ while filled circles designate
measurements of $\Dnuiss$. Arrows indicate upper limits.  The two quantities are related by
$2\pi\tau_d\Dnuiss = C_1$ with $C_1 = 1.16$ (see \S\ref{sec:expected} for
further details). All measurements have been scaled to 1 GHz from the
original frequencies assuming $\tau_d \propto \nu^{-4.4}$. The solid curve
is a least squares fit and the dashed lines represent $\pm 1.5 \sigma$
deviations from the fit.
\label{fig:plottau3}
}
\bigskip

\subsection{Dedispersion} \label{sec:dedisp}

While the effects of scattering cannot be removed with hardware or
software, the effects of interstellar dispersion can be largely or
completely corrected for. The degree to which this can be done depends on
the method of dedispersion used.  We discuss two methods of dedispersion
and compare the time resolutions achievable with each.

\subsubsection{Post-Detection Dedispersion}

Post-detection dedispersion operates on `detected' signals, or signals
which are intensity-like quantities because a squaring operation has taken
place after the signal has been channelized in a spectrometer.
Typically, these data are acquired using an analog or digital
multi-channel spectrometer, or `filterbank', in which the bandpass is
divided into a number of frequency channels and recorded at
short time intervals after detection. 
Post-detection dedispersion involves shifting the signal from
each frequency channel by the predicted dispersive delay for a particular DM. This
delay,
\be
t_{\rm DM} = 4.2\times10^{3} \, {\rm DM} \, \nu_{\rm GHz}^{-2} \, \mu s,
\label{eq:delay}
\ee
is consistent with the differential form of Eq.~\ref{eq:DtDM}.

The degree to which dispersion can be removed (i.e. the minimum time
resolution for this method of dedispersion)  is determined by the
competing effects of $\DtDM$ and $\DtDnu$, both of which depend on the
channel bandwidth. The optimal channel bandwidth corresponds to equality
of the terms $\DtDM$ and $\DtDnu$ and is therefore $\Dnu_{\rm MHz} =
\left( \nu_{\rm GHz}^3/8.3\rm DM\right)^{1/2}$. Assuming DM is known
exactly (i.e. $\dDM = 0$), the minimum time resolution achievable through
this method of dedispersion becomes
\be
\Delta t_0 &=&
       \left [
                2 ({\DtDM}_{\rm min})^2+ \DtISS^2
        \right ]^{1/2}
\ee
\rm\noindent where
\be
{\DtDM}_{\rm min} &\approx& \left ( 8.3 \DM \nu_{\rm GHz}^{-3} \right)^{1/2}
	\,\mu s
\label{eq:timeres2}
\ee
is the minimum dispersion smearing.

\subsubsection{Coherent Dedispersion}

Coherent, or pre-detection, dedispersion, operates on signals proportional
to the electric field selected by the receiver system, which contains
phase information. These data are typically acquired by bandpass filtering
signal voltages and shifting them to zero frequency (i.e. `baseband
mixing').  Baseband signals are sampled according to the Nyquist criterion
applied to the total bandwidth.  Coherent dedispersion is being used more
commonly as recording bandwidths and computer power have grown. It
involves removal of the dispersive phase rotation of the (complex)
electric field through a deconvolution procedure.  The function to be
deconvolved is approximately in the form of a `chirp' function having
duration equal to the dispersion time across the total bandwidth, given by
Eq.~\ref{eq:DtDM}.  This is often done using Fourier transforms (as
pioneered by Hankins in 1971) and has also been implemented using
finite-impulse response, time-domain filtering (e.g. Backer et al.
1997).
 If the source's DM is known exactly, then the phase
perturbation can be removed exactly and $\DtDM = 0$.  The minimum time
resolution achievable is then simply
\be
\Delta t_0  =
  \left [\DtDnu^2  + \DtISS^2
\right ]^{1/2}.
\ee

\subsubsection{Trial Dispersion Measures} \label{sec:trial}

In practice, the DM is not usually known in searches for radio transients.
Therefore, for either method of dedispersion, a range of trial DMs must be
used to dedisperse the data. The maximum DM searched can be estimated by
using a model for the Galactic electron density in the direction of the
source (e.g. Cordes \& Lazio 2002).  Because there are significant
model uncertainties owing
to unmodeled electron density structures in the ISM, 
it is wise  to
search to DM values higher than predicted by the model. The spacing of
trial DMs is determined by the maximum smearing one is willing to accept
in the final dedispersed time series.  This residual dispersion smearing
is given by $\DtdDM$, calculated using Eq.~\ref{eq:DtDM}-\ref{eq:DtdDM} with
$\Delta\nu_{\rm MHz}$ as the total bandwidth. The spacing of DM values can
be calculated by letting the error from $\DtdDM$ be a fraction $\epsilon$
of the total error calculated from Eq.~\ref{eq:timeres}. Given that
$\DtdDM =  (\dDM/\DM) \DtDM$ the required spacing of \DM\
values in the grid is
\be
\dDM /\DM  = \frac{\epsilon\Delta t} {\DtDM}
\ee
where, for optimal resolution, $\Delta t \to \Delta t_0$.
Note that as the DM increases, both $\DtDM$ and $\tau_d$ increase, making the optimal spacing between trial
DMs
greater at higher DMs.

We explore the effect that an incorrect DM has on an individual pulse's
amplitude. 
For a rectangular bandpass function and
for a Gaussian-shaped pulse with width $W$ (FWHM) in
milliseconds, the ratio of measured peak flux $S(\delta {\rm DM})$ to
true peak flux $S$ for a DM error $\dDM$ is
\be
\frac{S(\delta {\rm DM})}{S}
           = \frac{\sqrt\pi}{2}\zeta^{-1}{\rm erf}\,\zeta 
\ee
\rm where
\be
\zeta       = 6.91\times 10^{-3}\dDM 
		\,\frac{\Dnu_{\rm MHz}}{W_{\rm ms}\nu_{\rm GHz}^3}.
\ee
Figures~\ref{fig:dDMsnr} and \ref{fig:dDMsnr2} illustrate the dependence
of $S(\delta {\rm DM})/S$ on these quantities.   Figure~\ref{fig:dDMsnr}
implies a FWHM for $S(\delta {\rm DM})/S$ vs $\zeta$,
$\Delta\zeta \approx 3.5$, which corresponds to an  
 equivalent range in DM, 
\be
\Delta\DM \approx 506 \frac{W_{\rm ms} \nu_{\rm GHz}^3} {\Delta\nu_{\rm MHz}}.
\label{eq:dDM}
\ee

\medskip
\epsfxsize=9truecm
\epsfbox{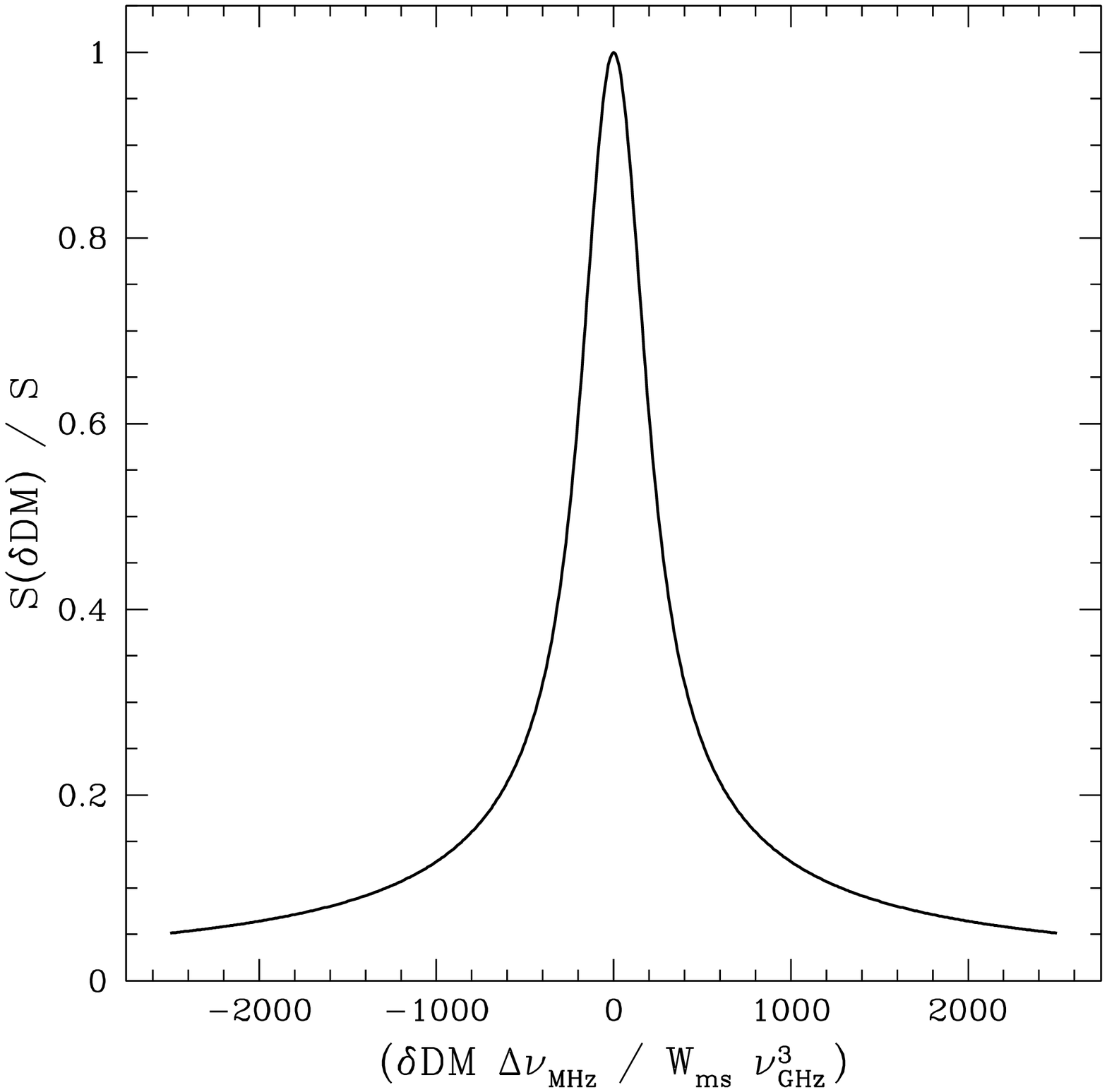}
\figcaption
{The ratio of measured peak flux to true peak flux as a function of $\dDM$,
$W$, $\nu_{\rm GHz}$ and $\Delta\nu_{\rm MHz}$ for $\dDM$ in pc cm$^{-3}$ and
$W$ in ms. The abcissa is a combined factor signifying that the
signal-to-noise degrades for larger errors in DM, smaller pulse widths,
lower frequencies and larger bandwidths.
\label{fig:dDMsnr}
}
\medskip
\epsfxsize=9truecm
\epsfbox{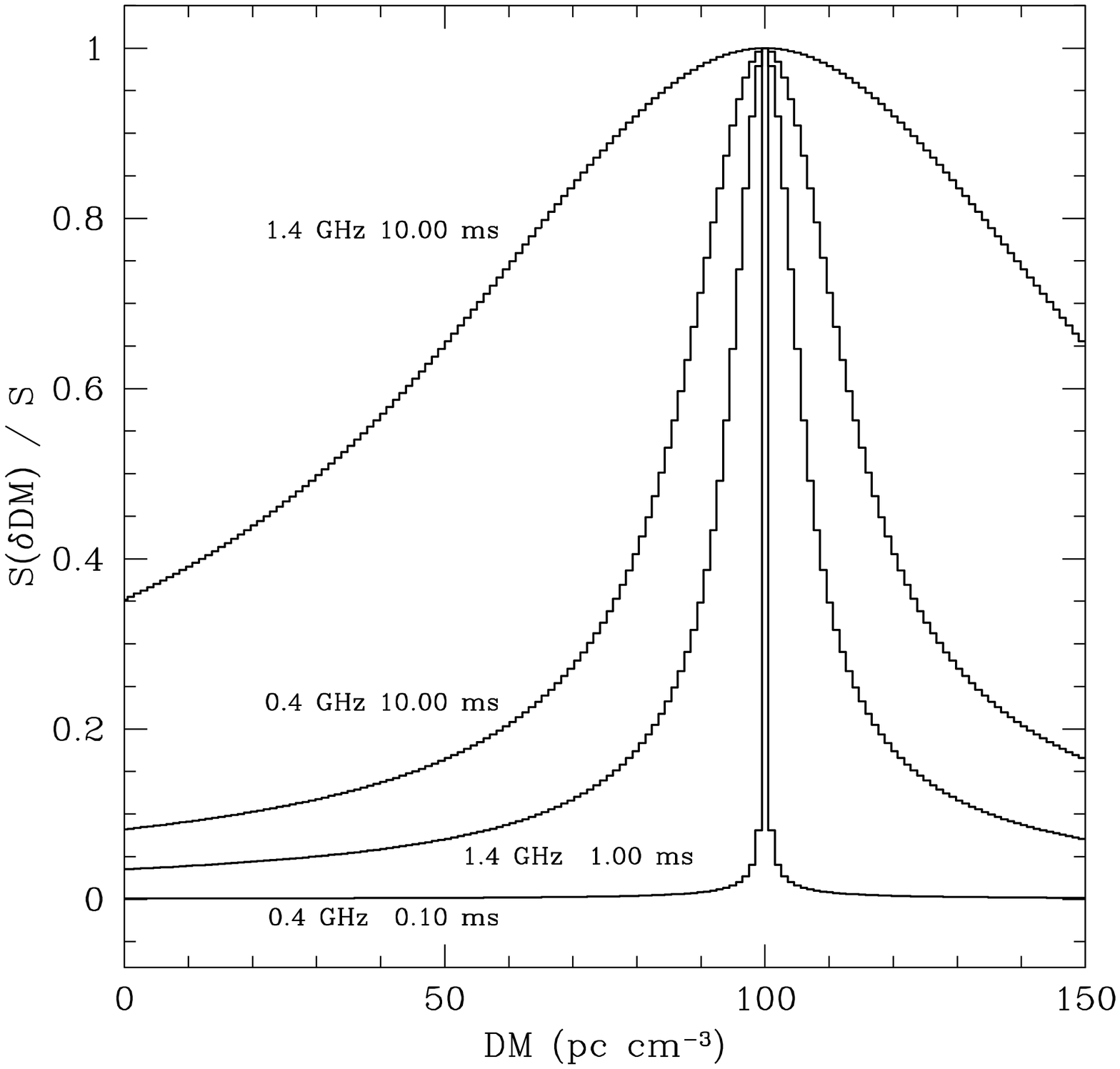}
\figcaption
{The ratio of measured flux to true flux as a function of trial DM used
for dedispersion for a pulse with true DM of $100$ pc cm$^{-3}$. Curves
are labeled with observing frequency and intrinsic pulse width. Bandwidths
of 10 and 100 MHz are assumed at 0.4 and 1.4~GHz, respectively.
\label{fig:dDMsnr2}
}

\subsection{Matched Filtering} \label{sec:match}

After dedispersion, each dedispersed time series, or `DM channel', must be
searched individually for pulses with amplitudes above some
signal-to-noise (S/N) threshold.  This search process is simply an
exercise in matched filtering, with the highest detected S/N achieved when
the effective sampling time of the time series is equal
to the detected width of the pulse. Since the intrinsic widths of signals
are typically unknown, as are the contributions to the width from
dispersion and scattering, a large parameter space must be searched. This
is most easily done by `smoothing' time series by adding adjacent samples
iteratively.  In the absence of knowledge about the true pulse shape and
width, the smoothing approach is a straight-forward and efficient
approximation to optimal detection.

To explore the S/Ns achievable through optimal matched filtering,
first consider a pulse whose intrinsic amplitude and width are $S_i$ and
$W_i$, respectively, in the absence of extrinsic broadening effects.  The
intrinsic pulse area $A_i \approx S_i W_i$.  Optimal detection implies a
signal to noise ratio
\be
(S/N)_i = \frac{1}{\sigma\sqrt{W_n}} \left (\frac{A_i}{\sqrt{W_i}}\right ),
\label{eq:snri}
\ee
where $\sigma$ is the rms noise in the time series (before any smoothing)
and $W_n$ is the correlation time for the noise.  
We assume that $W_i \gg W_n$. For the system noise $\ssys$ expressed as a 
flux density 
in Jansky units, the
rms radiometer noise is $\sigma = \ssys / (\npol\Delta\nu W_n)^{1/2}$,
where $\npol$ is the number of polarizations summed and $\Delta\nu$ is the
total bandwidth, so that S/N in the case of optimal detection is
independent of $W_n$.     
As can be seen, $S/N \propto A_i /\sqrt{W_i}$ so,
for fixed pulse area, a narrower pulse yields larger $S/N$. However, a
low-amplitude, broad pulse is more easily detectable than a sharp narrow
pulse if its area is sufficiently larger.

Now consider pulses that are broadened extrinsically by the interstellar
and instrumental effects discussed earlier and/or by pulse smearing from
orbital motion. 
All of these broadening effects conserve pulse area.
For heavily scattered pulses, the measured shape will be dominated
by the pulse broadening function, which is approximately 
a one-sided exponential in form.   The detailed shape is a function
of the wavenumber spectrum and spatial distribution of the 
scattering medium (Williamson 1972; Lambert \& Rickett 1999).
For such cases, the appropriate matched filter would be the 
shape of the pulse broadening function.
Pulse smearing from orbital motion
is negligible for known radio pulsar binaries (i.e.
orbital periods $\ge$ 1 hr) and for intrinsic pulse durations of seconds
or less.  However, for very short-period binaries, such as two neutron
stars undergoing inspiral from gravitational radiation, and for very
long-duration pulses, orbital motion may be important in single pulse
detection.  
If such broadening results in a net
pulse width $W_b$, then matched filtering yields
\be
(S/N)_b = (S/N)_i \left ( \frac{W_i}{W_b} \right)^{1/2}.
\label{eq:snrb}
\ee

\subsection{Thresholding}\label{sec:thresh}

To determine a reasonable threshold for a search for single pulses, we
assume that the statistics are Gaussian in the absence of any
signal, either celestial or RFI (a good assumption when the time-bandwidth
product $\gtrsim$~5). The expected number of candidate
pulses above threshold from radiometer noise alone  in a single 
DM channel in a time series of length $N$ is 
\begin{equation}
N(>{\rm threshold}) = N P_t  \sum_{j=0}^{j_{\rm max}} {2^{-j}}
	\approx 2N P_t, 
\label{eq:gauss}
\end{equation}
where $P_t$ is the integrated Gaussian probability of measuring a pulse above 
threshold, 
and $j_{max}$ is the
number of times a time series is smoothed and decimated.  
The approximate equality holds for $j_{\rm max}\gg 1$. 
The threshold can be chosen based on how many `false alarms' 
(i.e.  threshold crossings caused solely by radiometer noise) 
one is willing to accept. In reality, however, the measured statistics are not
Gaussian due to the presence of RFI and the level of RFI often determines
the threshold used.

Once pulses above threshold are recorded, the mean and rms can be
recalculated with these `first-pass' strongest pulses removed and the
search for pulses above threshold is repeated. This iterative procedure
keeps the mean and rms from being biased by a few very strong pulses. To
allow for any long-term level changes, small ($\sim$~seconds) subsets of
the original data set should be searched individually.

\subsection{Diagnostics}

Once all smoothed, dedispersed time series have been searched for pulses
above threshold, the output must be examined to determine whether these
pulses are of astrophysical origin. This can be done by creating
diagnostic plots like those shown in Figures~\ref{fig:crabgiant} and
\ref{fig:1937giant}, which show example search output\footnote{
  Similar plots for a number of pulsars can be found at
  {\tt http://www.jb.man.ac.uk/$\sim$mclaughl/single}
  and at
  {\tt http://www.astro.cornell.edu/$\sim$cordes. }
}
for the giant-pulse emitting pulsars PSR~B0531+21 and PSR~B1937+21. While the
examples here are from a search for single pulses from
pulsars, the analysis method is applicable to searches for any transient
radio signals.

The data for PSR~B0531+21 were taken at the Arecibo Observatory at a
frequency of 2330~MHz across a 100~MHz bandpass with the Wideband Arecibo
Pulsar Processor (WAPP, \verb+http://www.naic.edu/~wapp/+) with 64
frequency channels and 32 $\mu$s sampling. The data for PSR~B1937+21 were
taken at a frequency of 430~MHz across a 10-MHz bandpass with the Arecibo
Observatory Fourier Transform Machine (AOFTM,
\verb+http://www.naic.edu/~aoftm/+), which has 1024 frequency channels and
102.4 $\mu$s sampling. While the WAPP uses 16-bit sampling, the AOFTM uses
two-bit sampling, which will limit the measured dynamic
range of giant pulses.

\medskip
\epsfxsize=9truecm
\epsfbox{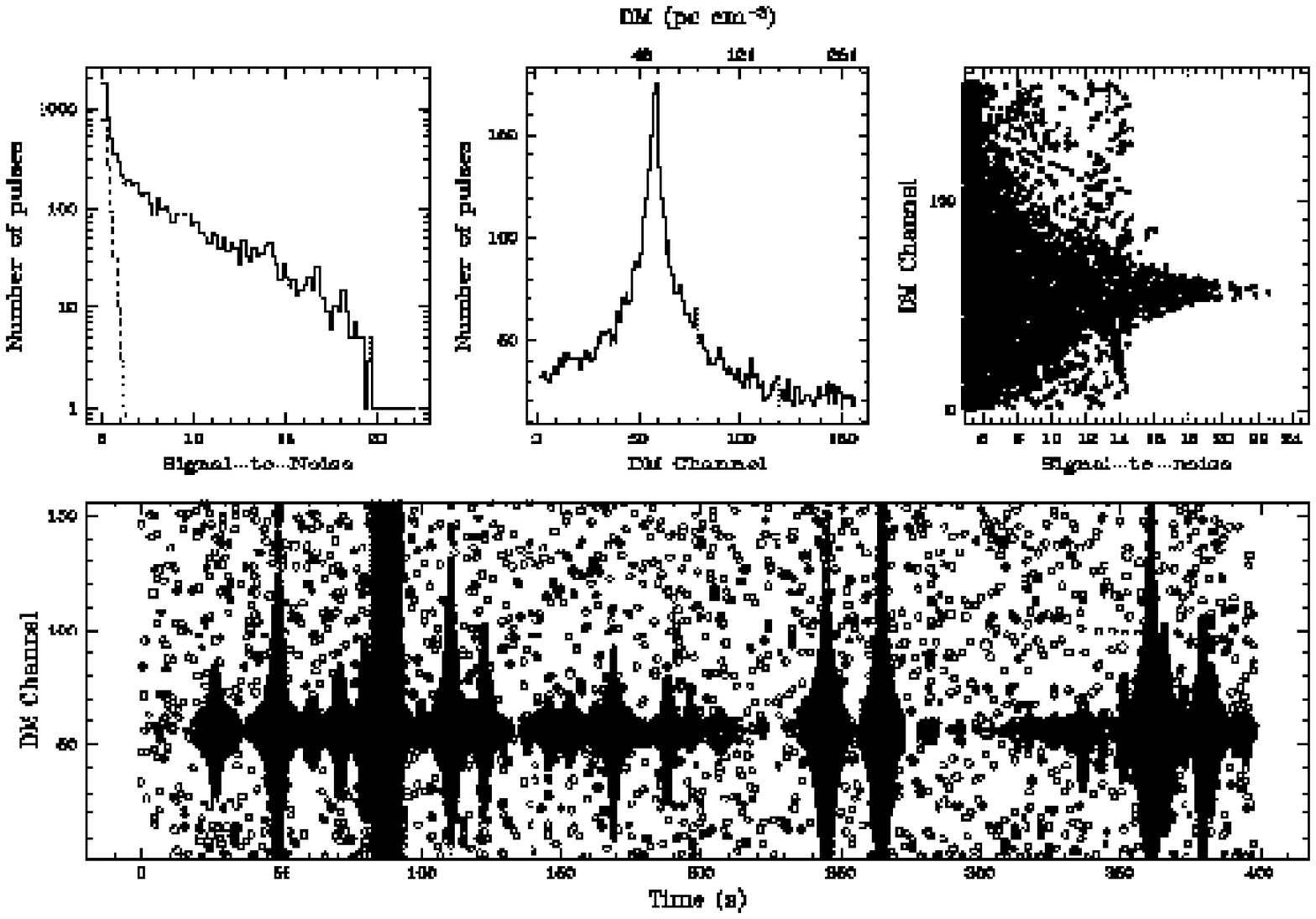}
\figcaption
{Single-pulse search results at 2330 MHz for the
Crab pulsar, whose DM is $56.8$ pc cm$^{-3}$.
The pulsar's period-averaged flux density $\sim$~3~mJy at this frequency.
As described in the text, each time series was
smoothed 7 times, corresponding to a maximum smoothing
$\sim$~13~ms. If a pulse is detected in more than one of the smoothed
time series, only the highest S/N result is recorded.
Most pulses were optimally detected
with smoothings of 2 or 3 samples, equivalent to widths of $\sim$~64 or
128~$\mu$s. Pulses that are strongest at $\DM = 0$ pc~cm$^{-3}$ are not included.
{\sc Upper left:} Histogram of S/N for identified pulses for S/N $> 5$
(solid line).  Note the logarithmic scale of the y-axis. The dashed line
shows the Gaussian distribution expected for noise only.
{\sc Upper middle:} Number of pulses above a 5~$\sigma$ threshold vs. DM
channel. The lower x-axis shows DM channel number, while the upper x-axis
shows DM in pc cm$^{-3}$.  A broadened peak at the DM of the Crab is obvious.
{\sc Upper right:} Scatter plot of DM and S/N.   The largest values of S/N occur
at or near the DM of the pulsar while the scatter of points
uniform over DM and extending to S/N $\sim 14$ is caused by
strong RFI at $\sim 90$ sec  in the time
series shown in the bottom frame.
{\sc Bottom:} All pulses
with S/N greater than 5~$\sigma$ plotted vs. DM channel and time. The size
of the circle is linearly proportional to the S/N of the pulse, with the
largest circle representing a S/N $\sim$~23.  As shown in
Figures~\ref{fig:dDMsnr} and \ref{fig:dDMsnr2}, strong pulses at high
frequencies may be detected across many DM channels.
\label{fig:crabgiant}
}

\bigskip

The data for both pulsars were dedispersed across a range of trial DMs
using the post-detection method as described in \S\ref{sec:dedisp}. As
discussed, DM channels are spaced more coarsely at higher DMs. A
single-pulse search was carried out as described in \S\ref{sec:match}. 
Figures~\ref{fig:crabgiant} and \ref{fig:1937giant}  each show
histograms of S/N and of DM,  a scatter plot
of S/N and DM,  and a time series of events above a threshold
(S/N $> 5$ and 4, respectively, for B0531+21 and B1937+21).
In the absence of any
signal (celestial or RFI), we expect the distribution of S/Ns to be
Gaussian.   For a time series with $N$ samples and $N_{\rm DM}$
trial values of \DM, the number of detections per bin of
size $\Delta$ in S/N is
\be
\Delta N_p \approx 2 N N_{\rm DM} \Delta f({\rm S/N}),
\ee
where $ f({\rm S/N})$ is a Gaussian probability density function
with unit variance.  The dashed lines in the upper left panels
of Figures~\ref{fig:crabgiant}-\ref{fig:1937giant} are plots
of $\Delta N_p$ for the two cases.

The S/N histograms for both pulsars clearly deviate from a
Gaussian distribution at high values of S/N.  
In the absence of signal, we
also expect the DM distribution of pulses to be flat, in the mean,
with fluctuations obeying Poisson statistics. 
The DM distributions clearly deviate from this form, showing well-defined
peaks at the dispersion measures of the pulsars.
From
Eq.~\ref{eq:gauss}, the expected number of pulses above
threshold due to radiometer noise alone $\approx$ 12 and 260 for the data in
Figures~\ref{fig:crabgiant} and \ref{fig:1937giant}, respectively. The
baseline (i.e. in DM channels other than that of the pulsar) number of
pulses above threshold for PSR~B0531+21 is higher than this number
because, as discussed in \S\ref{sec:trial}, strong, high-frequency pulses
may be detected across a broad range of DM channels. The baseline number
of pulses for the lower frequency observation of PSR~B1937+21 is similar
to that predicted for Gaussian noise, aside from a peak at zero DM due to
RFI. The lower halves of Figures~\ref{fig:crabgiant} and
\ref{fig:1937giant} show the strongest individual pulses plotted vs. DM
and time, with the size of the circle proportional to S/N. This plot is
complementary to the DM histogram; sources which emit many weak pulses may
be detected in the DM histogram but may not appear in the lower plot while
sources which emit only a few strong pulses may do the opposite.

\medskip
\epsfxsize=9truecm
\epsfbox{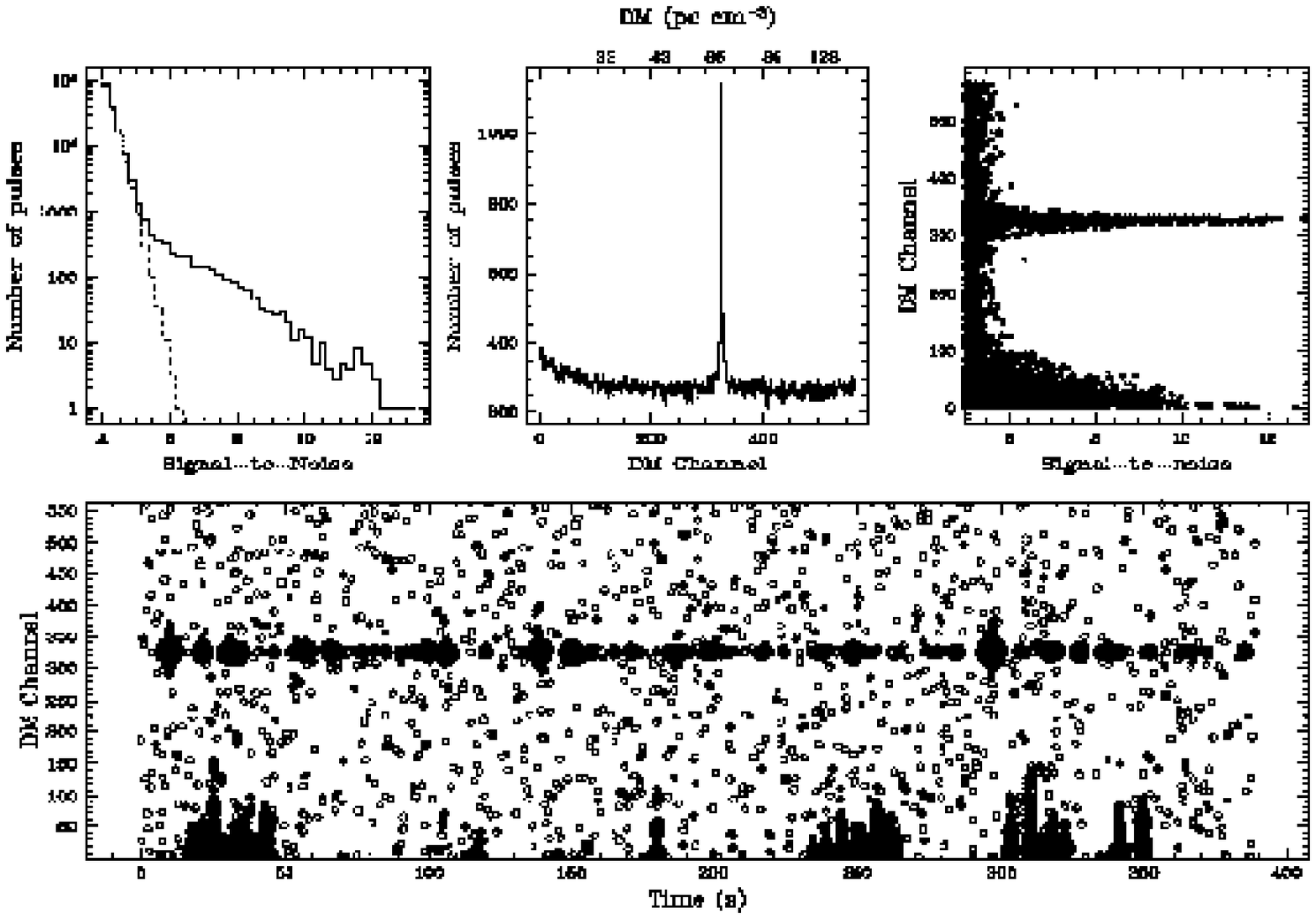}
\figcaption{
Single-pulse search results for PSR~B1937+21 at 430~MHz shown in
the same format as Figure~\ref{fig:crabgiant} and using a threshold of 4~$\sigma$.  This millisecond pulsar
has $\DM  = 71.04$ pc cm$^{-3}$ and period-averaged
430-MHz flux density $\sim$~200~mJy. Most pulses were detected
with no smoothing, equivalent to pulse widths $\lesssim 0.1$ ms.
The values of S/N are saturated because the data acquisition system
provides only 2-bit output samples, 1024 of which are summed in
the dedispersion procedure.
The narrowness of the histogram of DM values for detected pulses
(upper middle panel) is consistent with the narrow pulse widths
and Eq.~\ref{eq:dDM},
as demonstrated in Figure~\ref{fig:dDMsnr2}.
While there is RFI at low
DMs, an excess of strong pulses at the DM of the pulsar is obvious.
\label{fig:1937giant}
}
\bigskip

\section{Interstellar Scattering and Scintillation } \label{sec:iss}

In this section, we describe the basic scattering and scintillation
phenomena that are relevant to the detection and identification of transient
celestial signals.  The two main effects are pulse broadening and
intensity scintillations.  Both affect the detectability of transient
signals.  Pulse broadening always decreases S/N while scintillation can
increase or decrease S/N.  In addition, these effects can signify whether
a given candidate transient is truly celestial.

The degree of scattering is strongly frequency, direction, and distance
dependent.  Moreover, source sizes strongly influence the fractional
intensity modulation of scintillation.  Sources compact enough to emit
fast transients will generally show scintillations at centimeter (and
longer) wavelengths.  In the following, we describe quantitative aspects
of scattering and scintillation, observing regimes and optimal detection
strategies.

\subsection{Basic Phenomena}

Multipath scattering of radiation from electron-density variations in the
ISM produces the pulse-broadening effect discussed in \S\ref{sec:timeres}.
It also produces scintillations that modulate the measured intensity.
Diffractive interstellar scintillations (DISS), which occur in the
so-called strong scattering regime, produce intensity variations with
characteristic frequency scales $\Dnuiss \ll\nu$.  DISS is caused by
small-scale irregularities in the interstellar plasma and, for
observations at meter wavelengths, has typical time and frequency scales
of order minutes and   MHz, with large variations that depend
on DM, frequency and direction. Longer term, relatively broadband
intensity variations of timescales of days to months are caused by
refractive interstellar scintillations (RISS), associated with
larger-scale irregularities than those responsible for DISS. 
In the strong scattering regime, DISS and RISS are distinct 
phenomena. The intensity modulation from DISS, $g_{\rm d}$, is a random
variable with one-sided exponential statistics and 100\%
fractional variation for an intrinsically pointlike source. 
The modulation from RISS, $g_{\rm r}$, has a more symmetric distribution
with smaller rms variation of a few tens of percent. 
Reviews of the phenomena may be found in Rickett
(1990) and Narayan (1992). In weak scattering, DISS
and RISS merge to produce fractional intensity variations of less than
unity.  The transition from strong to weak scattering depends strongly on
direction. For lines of sight looking out of the Galactic plane, it is
typically $\sim$~5~GHz, but for the Galactic center the transition is
higher than 100 GHz.

\subsection{Source Size Requirements and Coherence}
\label{sec:iso}

Sources must be compact for DISS and RISS to occur.  In the strong
scattering regime, a point source will be broadened to a size $\thetad$
that is typically in the range of 1 mas to 1 arcsec at 1 GHz.  The
resultant diffraction pattern at an observer's location has spatial scale
$\ld \sim \lambda / \thetad$, where $\lambda$ is the wavelength.  By
reciprocity, sources with spatial extents larger than 
$\sim\ld$ will diminish the fractional
modulation of DISS (apart from `leveraging' effects associated with the
scattering region being closer to the source or to the observer, which we
discuss below).  We define the isoplanatic angular size as $\thetaiso
\propto \ld / D \propto \lambda / D\thetad$, where $D$ is the distance to
the source and the proportionality constant depends on the location and
depth of the scattering region along the line of sight. For RISS, the
isoplanatic scale $\lr \sim D\thetad$ and the isoplanatic angular size is
$\thetaisor \propto \thetad$. At 1 GHz, and for path lengths through the
ISM $\sim$~1~kpc, the isoplanatic angular scale $\thetaiso
\sim$~$10^{-7}$~arc~sec for DISS and $\sim 10^{-3}$~arc~sec for RISS,
though with large variations.

Transient sources producing short-duration signals are compact enough to
satisfy these criteria in some cases. Given the light-travel size for a
source at 1 kpc, signals shorter than 50~ms will show fully-modulated DISS
while signals shorter than 500 s will show fully-modulated RISS.
Relativistic motions, of course, will alter these estimates, by increasing
the light-travel source size by a 
Doppler factor $D_{\gamma} = 1/\gamma(1-\beta\cos\theta) \approx 2\gamma$
for $\gamma\gg 1$
(where $\gamma$
is the Lorentz factor of bulk motion in the source).  The transverse size
is relevant for scintillation quenching, however, and is related to the
light-travel size in a model-dependent way.

For longer path lengths or larger scattering strengths, the isoplanatic
angle for DISS becomes smaller while that for RISS becomes larger.  For an
extragalactic source, the greater distance increases the allowed signal
durations while, if viewed at Galactic latitudes $\gtrsim 10^{\circ}$ and
for similar host galaxy inclinations, the scattering angular size
($\thetad$)  remains small.

As is well known, fast transients necessarily arise from coherent emission
if the measured transient signature is intrinsic and implies a brightness
temperature in excess of physically plausible limits.  
Using the same light-travel size as before, the brightness temperature is
\be
T_b \approx 10^{14.8} \, K\, S_{\rm mJy}
	\left ( \frac{\dkpc}  {\nu_{\rm GHz}\Delta t} \right)^2;
\label{eq:tb}
\ee
$S_{\rm mJy}$ is the flux density in mJy and $\Delta t$ is the signal duration.
Relativistic motion decreases the coefficient by a factor
$\sim 1/4\gamma^2$ for $\gamma\gg 1$.  Sources compact enough to show DISS 
(i.e.  $\Delta t \lesssim 50$ ms for $\gamma\sim 1$) have $T_b \gtrsim
10^{17.4}\,K$ for other parameters held fixed in Eq.~\ref{eq:tb}. This is
too large to be associated with incoherent processes.

\subsection{Expected Timescales and Bandwidths}
\label{sec:expected}

Both DISS and pulse broadening are well established for Galactic pulsars.
RISS has been identified in observations of AGNs and masers as well as
pulsars.  For a given combination of line of sight and frequency, DISS and
pulse broadening are typically mutually exclusive in pulsars, because when
the pulse broadening is large enough to be identified, the scintillation
bandwidth is too small to be measured, as indicated by values
plotted in Figure~\ref{fig:plottau3}. Exceptions occur, of course,
because the identification of pulse broadening is a function of intrinsic
pulse width.

Quantitatively, the pulse broadening time $\tau_d$ and the scintillation
bandwidth $\Dnuiss$ are reciprocally related by $2\pi\tau_d\Dnuiss = C_1$
(e.g. Cordes \& Rickett 1998).  For a uniform medium with Kolmogorov
fluctuations, $C_1 = 1.16$. Otherwise, $C_1$ $\sim$~1, except for
scattering in thin screens near the source or observer. For typical pulsar
pulses, pulse broadening is manifest when $\tau_d \gtrsim 1$~ms, in which
case the scintillation bandwidth is too small to be resolved (for typical
source flux densities and spectrometer resolutions).  Exceptions to this
are the Crab pulsar's giant pulses, which show structure in the range of a
few nanoseconds to a few microseconds (Hankins 2003).
These narrow pulses allow measurement of much smaller pulse broadening
times, corresponding to measurable scintillation bandwidths.
If other sources of very narrow pulses exist (e.g. $W\ll 1$ ms),
then scintillation frequency structure will be measureable for them
as well.

\medskip
\epsfxsize=9truecm
\epsfbox{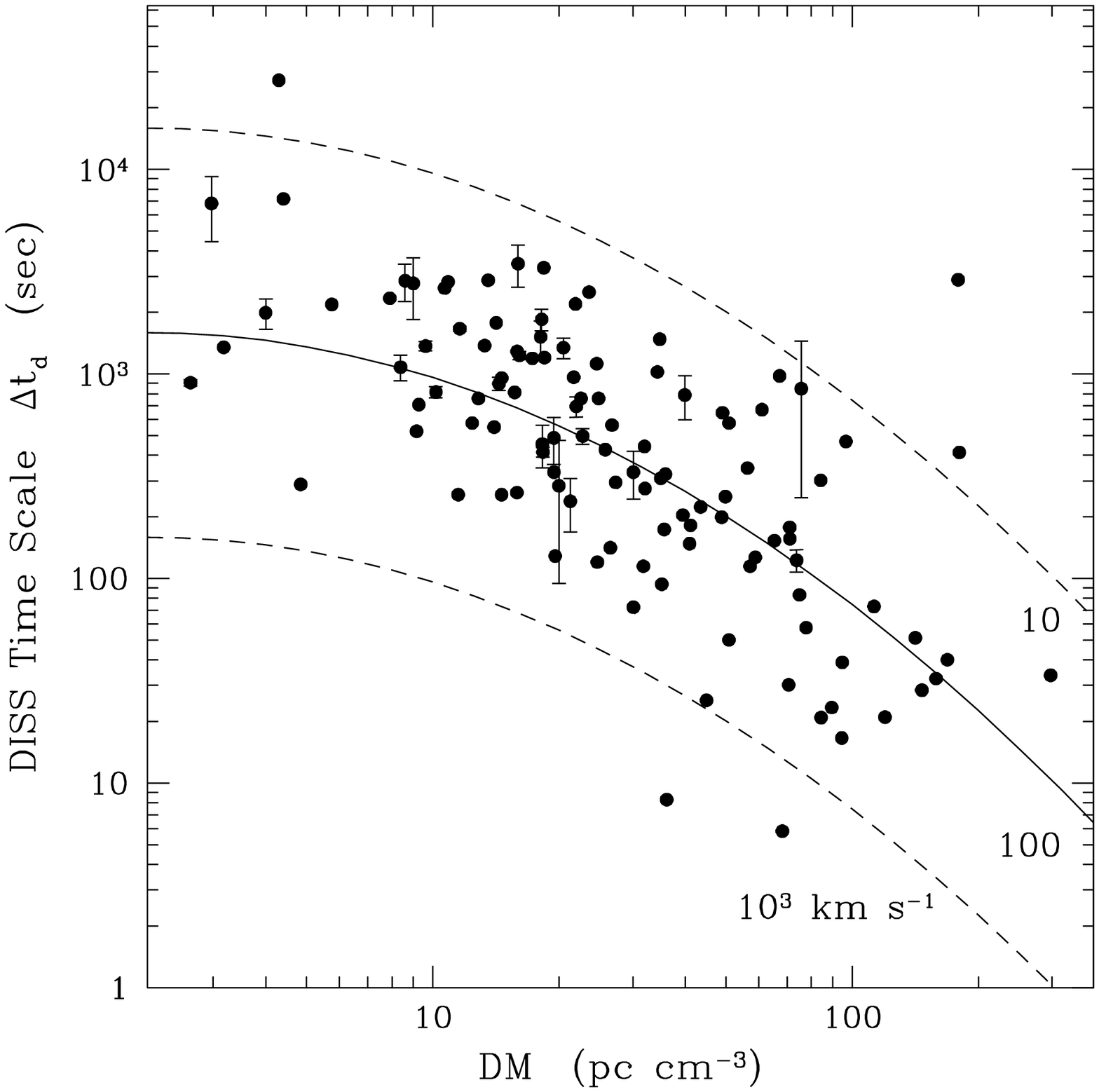}
\figcaption{
The characteristic DISS timescale at 1 GHz plotted against DM for a sample
of pulsars. For objects with multiple observations, the vertical bars
designate the $\pm 1 \sigma$ variation of the mean value.  The solid and
dashed lines show predicted DISS timescales for transverse source
velocities of 10, 100 and $10^3$ km s$^{-1}$ using Eqs.~\ref{eq:taud} and
\ref{eq:dtiss2} and the relation $2\pi\tau_d\Dnuiss = C_1$ (see text).
For these curves, we also assume the scattering medium is statisically
uniform, which is approximate at best, and we do not include the scatter
in values about Eq.~\ref{eq:taud} evident in Fig~\ref{fig:plottau3}.
The DISS timescale varies with frequency as $\Dtiss \propto \nu^{1.2}$ if the
scintillation bandwidth has the Kolmogorov scaling, $\Dnuiss \propto
\nu^{4.4}$, as appears consistent for some objects.
For extragalactic sources, the DISS time scale will differ because the
geometry of the scattering medium consists of a foreground region from
the Galaxy and another region corresponding to material in the host
galaxy (if any).  Nonetheless,  the order of magnitiude value of
the time scale can be estimated using the value of DM expected from
just the foreground material in the Galaxy (see Appendix~\ref{app:iss}).
The points are from
Bhat et al. (1999); Bogdanov et al. (2002); Camilo \& Nice (1995);
Cordes (1986); Dewey et al. (1988); Foster et al. (1991); Fruchter et al.
(1988); Gothoskar \& Gupta (2000); Johnston et al. (1998); NiCastro et al.
(2001); and Phillips \& Clegg (1992).
\label{fig:dtvsDM}
}
\bigskip

While the pulse broadening timescale and DISS bandwidth depend on the
distribution of scattering material along the line of sight, the DISS
timescale $\Dtiss$ also depends on the source's velocity. The
scintillation velocity, $\viss$, is the speed of the ISS diffraction
pattern across the LOS and determines the scintillation timescale as
\be
\Dtiss \equiv \frac{\ld}{\viss} = W_1 \frac{\ld}{\vsperp}
 = \frac{W_2}{\nu\vsperp} \left( \frac{cD\Dnuiss}{4\pi C_1} \right)^{1/2},
\label{eq:dtiss}
\ee
where $\vsperp$ is the source velocity and $W_{1,2}$ are weighting factors
that take into account the location of scattering material along the LOS
(c.f. Cordes \& Rickett 1998, Eq.~4, 9, 17 and 23). The final equality in
Eq.~\ref{eq:dtiss} follows by relating the spatial scale $\ld$ to the
scintillation bandwidth $\Dnuiss$ and other quantities  
by using $\ld = \lambda / \thetad$ (c.f. \S\ref{sec:iso}) along with
the proportionality $\taud \propto D\thetad^2  / 2c$ and the
equation relating $\taud$ to $\Dnuiss$.
For nominal
values, and assuming a Kolmogorov spectrum for the scattering irregularities, the scintillation timescale is
\be
\Dtiss = 252 \,s\,
	W_2
	(cD_{\rm kpc}{\Dnuiss}_{\rm,MHz})^{1/2}
	(\nu_{\rm GHz}{\vsperp}_{,100})^{-1},
\label{eq:dtiss2}
\ee
where ${\vsperp}_{,100} = \vsperp / (100 \,{\rm km\, s^{-1}})$ and $W_2 = 1$ for a uniform medium.
 Pulsars
have 3D velocities that average $\sim 400$ km s$^{-1}$ (e.g. Cordes \&
Chernoff 1998), so their scintillation timescales can be quite short.
Other Galactic sources may be slower and would accordingly have longer
scintillation timescales.  Note that since $\Dnuiss \propto \nu^{4.4}$ for
a Kolmogorov wavenumber spectrum for the electron-density irregularities,
$\Dtiss\propto \nu^{1.2}$ in the strong-scattering regime.  In weak
scattering the characteristic scintillation bandwidth $\approx \nu$ and
the level of modulation becomes less.
Figure~\ref{fig:dtvsDM} shows the scintillation time scale for
known pulsars along with estimates based on Eq.~\ref{eq:dtiss2}
for transverse velocities of 10, 100 and $10^3$ km s$^{-1}$.

\subsection{Comparison of Galactic and Extragalactic Source Scintillations}

In some circumstances, the scintillations of fast radio transient sources
can serve as a `reality check' on the astrophysical nature of the
emission, based on the direction, distance and frequency of observation.
In order to estimate the DISS and pulse broadening properties of
extragalactic transient sources, we have developed the formalism in the
Appendix. As shown there, we expect scintillation timescales
for extragalactic sources to be about one-third as much as for Galactic
sources in the same direction. We also find that the pulse broadening time
for an extragalactic source will be approximately six times larger than
for a Galactic source in the same direction. Similarly, scintillation
bandwidths for extragalactic sources will be about six times smaller.

\subsection{Optimal Observing Strategy in the Strong Scattering Regime} \label{sec:issopt}

DISS in the strong scattering regime can both help and hinder detection of
transient radio signals.  Signals too weak to be detected without the
scattering medium may be modulated above the detection threshold while
stronger signals that are nominally above the threshold can be modulated
below.  As shown in Cordes \& Lazio (1991), in the strong scattering
regime (i.e. distances $\ge$ 1 kpc at $\sim$ 1~GHz), multiple observations of a
given target comprise a strategy that can be superior to single
observations even when the total time per target is held fixed.  The
requirements for favoring this strategy are that (a) the source size is
much smaller than the isoplanatic size and (b) that there is no bandwidth
averaging of the DISS frequency structure.

We now define several observational regimes.  In radio transient searches,
it is useful to consider the dynamic spectrum (intensity as a function of
time and frequency) as the analyzable data unit.  We follow Cordes \&
Lazio (1991), who define the number of independent scintillation
fluctuations (`scintles') contained in a frequency-time resolution element
of size $\Delta\nu \times \Delta t$ as
\be
\Niss &=& N_t N_{\nu}, \\
N_t &\approx& 1 + \eta \Delta t / \Dtiss, \\
N_{\nu} &\approx& 1 + \eta \Delta \nu / \Dnuiss,
\ee
where $\eta \approx 0.1-0.2$ is a `filling' constant. 
Given the large variation in scintillation parameters (c.f.
Figure~\ref{fig:plottau3} and Eq.~\ref{eq:dtiss2}), 
$\Niss$ varies by orders of magnitude 
as a function of frequency and DM.
For fast transients,
we expect $\Delta t \ll \Dtiss$, so that $N_t = 1$.  However, if
differential dispersive arrival times span a long time, which will be true
for large \DM\ or low frequencies, the dedispersed signal may span one or
more independent `scintles' and $N_t > 1$.  For broadband transients,
optimal detection requires integration over frequency (with implied
dedispersion), in which case it may also be true that $N_{\nu} \gg 1$.
Narrowband transients, such as may be expected from masers and from
signals from extraterrestrial intelligence, are likely to have $N_{\nu} =
1$, except for long lines of sight through the inner Galaxy combined with
observations at low frequencies, for which $\Dnuiss$ can be less than 1
Hz.

Three scintillation regimes may be identified in terms of $\Niss$. We
assume that the strong scattering regime applies and that the source size
is much smaller than the isoplanatic angular size:
\begin{enumerate}
\item $\Niss = 1$:  The measured signal will be 100\% modulated
and the intensities will be exponentially distributed.
\item $1 \lesssim \Niss \lesssim 100$:   The scintillations will
be less than 100\% modulated and will have a $\chi^2$ distribution
with $\sim 2\Niss$ degrees of freedom.  For $\Niss \gtrsim 5$, the distribution
is well described as a Gaussian function.   Fast transients with
$N_t = 1$ in this regime have multiple scintles across the frequency
range $\Delta\nu$.   Each scintle has an amplitude described by
the exponential distribution.  Before integrating over frequency,
the optimal S/N may result by weighting the individual amplitudes to favor
the strong scintles.
\item $\Niss \gg 100$:  The DISS modulation of the combined signal
(after integrating over time and frequency) $\lesssim 10$\%.
\end{enumerate}

In the first regime, where the signal is fully modulated, an
optimization strategy that maximizes the detectability of a source
with respect to DISS comprises multiple observations.
In a single observation, it is more likely that the
signal is modulated below its mean (probability = $1-e^{-1}$) than above
it.    Less frequently, a large boost in signal strength
is expected because the modulation has an exponential distribution.
If the total amount of telescope time is fixed and divided
into $N$ subobservations separated by more than a characteristic
scintillation time, there is an optimal, finite value of $N$. 
For an {\it intrinsically steady} source, the optimal number of
trials $\sim 3$ maximizes the probability that {\it at least} one
detection is made.  For transient sources, the optimization also depends
on the (unknown) intrinsic distribution of emitted amplitudes.  For
infrequent or single-instance transients, DISS in the first regime will on
average degrade the detectability of a given source.  However, across a
population of sources the DISS will aid detection in $\sim e^{-1}$ of the
cases and hinder detection in the remainder.

\section{Radio Frequency Interference}\label{sec:RFI}

Radio frequency interference is a growing problem for radio astronomy
in general and for the detection of transient signals in particular.
Considerable work is now being done on identifying and mitigating RFI
in a number of ways.   A situation similar to pulse detection in the
time domain is the detection of narrow spectral lines.  As discussed
by Horowitz \& Sagan (1993), RFI from artificial, terrestrial sources
can be identified, in part, by requiring candidate
signals to conform to  what is expected from celestial signals.
For narrow spectral lines,  one expects variable Doppler shifts in accordance
with Earth's spin and orbital motion.   As shown by Horowitz \& Sagan, 
however,
even this requirement allows rare RFI to satisfy basic requirements
that a real, celestial signal must satisfy.   For time domain
searches, such as those we have focused on here,  RFI can mimic
the dispersive arrival times of pulses in instances where the RFI
is a swept-frequency signal.    Thus, while filters based on
anticipated properties of celestial signals are necessary elements
of transient searches, they are not sufficient for providing 
candidate detections that are devoid of false positive detecions.

To identify real celestial transients while rejecting 
RFI, searches using single pixel systems must rely on 
application of tests for properties that real, celestial signals
must have, including   Doppler shifts and  dispersive arrival times,
and also any repeatability of the transients that can be explored
through multiple observations and on-and-off source comparisons.
Scintillations play an important role in tests for repeatability,
especially for compact sources that underly fast transients,
as discussed in Cordes, Lazio \& Sagan (1997).


\section{Conclusions} \label{sec:conclusions}

We have described issues related to and methods for searching for fast
radio transients. A very small parameter space of the transient radio sky
has been explored thus far. While we have gained an excellent understanding
of the high-energy transient sky through X-ray and gamma ray satellites,
radio searches for transient sources have been limited by the small sky
coverage and high susceptibility to RFI of current telescopes. To increase
the capabilities of single-pulse searches, systems with multiple beams or
multiple antennas are necessary in order to perform coincidence tests
which will discriminate between pulses of astrophysical and terrestrial
origins.  Radio telescopes with larger fields of view are necessary to
maximize the chances of detecting radio-bursting objects and to complement
the searches at high-energies. Larger collecting areas are needed to
increase our sensitivity to pulses from pulsars and other transient
sources in other galaxies. Multi-beam systems for the Arecibo and Green
Bank telescopes will likely be built within the next few years. Observing
with multiple beams will allow us to distinguish RFI from real signals. On
a longer time scale, large radio arrays such as the Square Kilometer Array
(SKA) will be developed. With a planned field of view over 100 times that
of Arecibo and a sensitivity over 20 times that of Arecibo, the SKA will
revolutionize our understanding of the transient radio sky. While the SKA
will not be operational for many years, single-pulse searches in the
meantime, especially with multiple beams, are essential for optimizing
routines for RFI excision and for processing large amounts of data. With
more sensitive instruments and more sophisticated analyses, the next few
decades will undoubtedly bring about a greater understanding of
radio-bursting objects, including Crab-like pulsars in other galaxies,
counterparts to high-energy bursting sources, and other classes of objects
which are yet to be discovered.

We thank Joe Lazio and Ira Wasserman for useful discussions.
This work was supported by NSF grants AST9819931, 
AST0138263 and  AST0206036  to Cornell University.  
MAM is also supported by an 
NSF Math and Physical Sciences Distinguished
Research Fellowship. Arecibo Observatory is operated by the National
Astronomy and Ionosphere Center, which is operated by Cornell University
under cooperative agreement with the National Science Foundation (NSF).
The Parkes Observatory is part of the Australia Telescope which is funded
by the Commonwealth of Australia for operation as a National Facility
managed by CSIRO.

{}

\clearpage
\appendix

\section{Interstellar Scattering and Scintillation
 of Sources in Other
Galaxies} \label{app:iss}

Scattering of radio waves from electron-density variations produces
angular blurring (or `seeing') of radio sources and (for variable sources)
temporal broadening due to multipath propagation.  Under certain
conditions, radiation from a source of spatially coherent emission will
show scintillation from interference between the differentially-arriving
wavefronts and from refractive changes in the radius of curvature of the
wavefronts. These various effects are associated with different
line-of-sight (LOS) weightings of the scattering material.  Here we take
these weightings into account when deriving estimates for scattering and
scintillations of extragalactic sources.

\newcommand{\cnsq}{C_n^2}
\newcommand{\veffperpvec}{{\bf V_{\rm eff,\perp}}}
\newcommand{\dxperpvec}{{ \mbox{\boldmath $\delta x_{\perp}$} }}
\newcommand{\xvec}{{\bf x}}
\newcommand{\dxvec}{{ \mbox{\boldmath $\delta x$} }}
\newcommand{\veffvec}{{\bf V_{\rm eff}}}
\newcommand{\vpvec}{{\bf V_{\rm s}}}
\newcommand{\vobsvec}{{\bf V_{\rm obs}}}
\newcommand{\vismvec}{{\bf V_{\rm m}}}
\newcommand{\dphi}{D_{\phi}}
\newcommand{\dt}{\delta t}
\newcommand{\dne}{{\delta n_e}}
\newcommand{\FWHM}{\theta_{\rm FWHM}}
\newcommand{\vpperpvec}{{\bf   {V_{\rm s}}_\perp}}
\newcommand{\vobsperpvec}{{\bf {V_{\rm obs}}_\perp}}
\newcommand{\vismperpvec}{{\bf {V_{\rm m}}_\perp}}
\newcommand{\nutrans}{\nu_{\rm trans}}
\newcommand{\Deff}{D_{\rm eff}}

The phase structure function may be used to calculate the angular
scattering and scintillation timescale. This function, as given by Cordes
\& Rickett (1998; Eq. 2), is
\begin{equation}
\dphi(\dxvec,\dt) = (\lambda r_e)^2  f_{\alpha} \int_0^D ds\, \cnsq(s)\,
       \left \vert \veffperpvec\dt+ \left(\frac{s}{D}\right)\dxperpvec
       \right \vert^{\alpha},
\label{eq:sf}
\end{equation}
assuming that diffraction involves only small angles and that the
underlying structure function for the electron density is isotropic and
scales as $\vert\dxvec\vert^{\alpha}$, where $\alpha+2$ is the power law
index of the electron density wavenumber spectrum.  For a Kolmogorov
wavenumber spectrum, $\alpha = 5/3$, subject to constraints on the
wavenumber cutoffs. In Eq.~\ref{eq:sf}, $\lambda$ is the wavelength, $r_e$
is the classical electron radius, $f_{\alpha}$ is a constant of order
unity which depends on $\alpha$, $\veffvec$ is the effective velocity of
the source, and $D$ is the distance to the
source. The integral is calculated from $s=0$ at the source to $s=D$ at
the observer's location and $\veffvec$ is
\begin{equation}
\veffvec(s) = \left (1-\frac{s}{D}\right )\vpvec + 
	\left(\frac{s}{D}\right)\vobsvec - \vismvec(s),
\label{eq:veffvec}
\end{equation}
where $\vpvec$, $\vobsvec$ and $\vismvec$ are the velocities of the
source, observer and medium (all relative to a common standard of rest).
Only the components perpendicular to the LOS affect the time
scales of diffractive interstellar scintillation (DISS).

In the following we assume that the scattering is `strong,' in the sense
that the phase perturbation induced by scattering irregularities
is larger than one radian when one considers two ray paths separated
by a Fresnel scale.  For a single thin screen, this scale is well
defined, whereas for a distributed scattering medium one must weight
the scattering material according to a factor determined by where it
lies along the path. For further details, see Rickett (1990).

\subsection{Angular Broadening}

First we consider angular broadening (i.e. ``interstellar seeing'').  The
observed angular diameter of a source is related inversely to the
transverse scale on which $\dphi(\dxvec, 0) = 2$.  The integral for
$\dphi$ weights scattering material by $(s/D)^{5/3}$ (for a Kolmogorov
spectrum) and thus favors material nearest the observer.  For objects in
other galaxies, we therefore ignore contributions to $\dphi$ from the host
galaxy. Consider an extragalactic source at distance D and Galactic
latitude $b$ and a Galactic source at the same latitude but at distance
$D_{\rm g}$. Introducing the scattering measure, defined as $\SM =
\int_0^D ds\, C_{n}^{2}(s)$, the ratio of angular diameters of these
sources as seen by an observer in the Milky Way is
\be
\frac {\FWHM({\rm xgal})}{\FWHM({\rm Gal})} =
		\left (
		\frac{\SM_{\rm Gal}}
		     {\int_{\rm 0}^{\rm D_{\rm g}} ds\, \cnsq(s) (1-s/D_{\rm s})^{\alpha}}
		\right )^{1/\alpha} \ge 1,
\label{eq:diamratio}
\ee
where $\SM_{\rm Gal}$ is the Galactic contribution to $\SM$ in the
direction to the source. The integral in the denominator is over the line
of sight to the source, with the integration now going from observer (at
$s=0$) to source.  Assuming an exponential distribution for $\cnsq$ and
considering a Galactic source at distance $D_{\rm g}$ equal to one
exponential scale height H, we find ${\FWHM({\rm xgal})}/{\FWHM({\rm
Gal})} \approx 2.1 $.  This approximate doubling of the angular diameter
of an extragalactic source viewed through the same medium as a Galactic
source results from the much smaller curvature of the extragalactic
source's wavefronts impinging on the ISM and can serve as a potential
distance estimator for high-latitude transient sources.

\subsection{Pulse Broadening}

As discussed in Cordes \& Lazio (2002), 
the pulse broadening timescale is proportional to
$\SM_{\tau}^{6/5}D$, where
\be
\SM_{\tau} = 6\int_0^D ds\, (s/D)(1-s/D) \cnsq(s).
\ee
In contrast to angular broadening, the LOS weighting function
is symmetric about the LOS midpoint, so a significant contribution is
expected from the host galaxy as well as trom the Milky Way.  
Consider a pulsed source in the host galaxy and a path length 
through the host galaxy $D_{\rm xgal}$  and a path length through 
the Milky Way of $D_{\rm g}$. Both $D_{\rm xgal}$ and $D_{\rm g}$ 
are much less than the total distance, $D$. For simplicity, we assume
that the scattering media in the two galaxies are uniform and that the
intergalactic medium makes no contribution to the scattering.
The ratio of the pulse-broadening time for the extragalactic object 
to that for a Galactic source at distance $D_{\rm g}$ 
from the observer is
\be
\frac{{\tau_d}_{\rm xgal}} {{\tau_d}_{\rm Gal}} \approx 3.7
	\left (
	1 +  \frac{{\SM}_{\rm xgal} D_{\rm xgal} } { {\SM}_{\rm Gal} D_{\rm g} }
	\right )^{6/5}
	\left (
	  \frac{D_{\rm g}}{D}
	\right )^{1/5}.
\label{eq:tau.ratio}
\ee
The particular form (with the exponent of 6/5 and the ratio of distances
to the 1/5 power) follows from the assumption
that scattering is dominated by
the power law portion of the wavenumber spectrum and not the
inner scale of the spectrum.   For intense scattering at low
frequencies, the inner scale may play a role, in which case
$6/5\to 1$ and the ratio of distances  is replaced by unity (see Cordes \& Lazio 2002
for details).
The factor of 3.7 in Eq.~\ref{eq:tau.ratio} results from the fact that the
Galaxy's ISM sees spherical waves from Galactic sources but plane waves
from extragalactic sources. Similarly, the host galaxy's ISM sees spherical 
waves from the emitting source.  
When the second term in square brackets is unity (equal contributions
from the host galaxy and the Milky Way), the coefficient of the factor
$ ( D_{\rm g} / D )^{1/5}$ becomes 8.5.   For a pulsar
in the Large Magellanic Cloud ($D \approx 50$ kpc), 
for example, the net ratio is about 3.9 for $D_{\rm g} = 1$ kpc. 
For an object in M33 ($D\approx 900$ kpc), the ratio is about 2.2.
Accordingly, the scintillation bandwidths are a factor of two to four
smaller.  Of course, scattering in the host galaxy could be
much larger than the Galactic scattering, yielding much larger
$\tau_d$ and much smaller $\Delta\nu_d$.  Representative
values of ${\tau_d}_{\rm Gal}$ can be read off of Figure~\ref{fig:plottau3}
by approximating $\DM \approx 30/\sin \vert b \vert$ pc cm$^{-3}$.

\subsection{Scintillation Time Scale}

We calculate $\Dtiss$, the scintillation timescale for DISS, using
$\dphi(0, \dt) = 1 $ to define the $1/e$ scale of the intensity
correlation function.  The LOS weightings in
Eq.~\ref{eq:sf}-\ref{eq:veffvec} imply that contributions from both the
host Galaxy and the Milky Way are important.  With $s/D\ll 1$ for the host
Galaxy, $\veffvec(s) \approx \vpvec - \vismvec(s)$.  For the portion of
the LOS within the Milky Way, $s/D\to 1$ and $\veffvec(s) = \vobsvec -
\vismvec(s)$. Evaluating $\dphi$ using the separate contributions to \SM\
from the host galaxy ($\SM_{\rm xgal}$) and the Milky Way ($\SM_{\rm
Gal}$), we find that
\be
\Dtiss =
	\left [
	  (\lambda r_e)^2 f_{\alpha}
	  \left(
		\SM_{\rm xgal}
		\vert
			\vpperpvec - {\vismperpvec}_{,\rm xgal}
		\vert^{\alpha}
		+
		\SM_{\rm Gal}
		\vert \vobsperpvec - {\vismperpvec}_{\rm ,Gal }
		\vert^{\alpha}
	  \right )
	\right ]^{-1/\alpha}
\label{eq:dtiss3}
\ee
Eq.~\ref{eq:dtiss3} implies that if the Galactic and extragalactic
contributions to the \SM\ are similar, both ends of the LOS contribute
equally to $\Dtiss$. However if the source velocity exceeds the velocities
of the observer and the interstellar media, as is typically the case with
pulsars, then the contribution from the host galaxy will dominate the
scintillation timescale.  For this case, we calculate the ratio of
$\Dtiss$ for an extragalactic source to that for a Galactic source as
\be
\frac{\Dtiss({\rm xgal})} {\Dtiss({\rm Gal})} \approx
	\frac{\vert \vpperpvec({\rm Gal}) \vert} {\vert \vpperpvec({\rm xgal})}
	\left [
	  \frac{\SM_{\rm Gal}}{\SM_{\rm xgal}}
	  \int_0^1 dx\, e^{-x} x^{\alpha}
	\right ]^{1/\alpha}
	\approx
	0.36
	\frac{\vert \vpperpvec({\rm Gal}) \vert} {\vert \vpperpvec({\rm xgal})}
	\left [
	  \frac{\SM_{\rm Gal}}{\SM_{\rm xgal}}
	\right ]^{1/\alpha},
\label{eq:dtissratio}
\ee
again assuming an exponential distribution for $\cnsq$ with a galactic
source at a
distance of one exponential scale height. Thus, along the same direction,
we expect the scintillation timescale of an extragalactic source to be
about 1/3 that of a Galactic source at a distance of one scale height for
$\cnsq$. This result is more general than 
the analysis of Cordes \& Rickett (1998),
who considered the scintillation time for extragalactic sources under the
assumption that the transverse velocity of the extragalactic source was
negligible and that there was no contribution to \SM\ from the host
galaxy.

\subsection{Weak Scintillation}

When scattering is weak,  the scintillations of a point source
have a modulation index satisfying $\sigma_I/\langle I \rangle <<1$,
the scintillations are correlated over a frequency range
comparable to the (center) radio frequency, and any pulse broadening
becomes smaller than the reciprocal radio frequency.   The apparent
source structure includes, in this regime, an unscattered component
along with a scattered component.   If $\sigma_{\phi}\ll 1$ is the
rms phase variation between two paths separated by a Fresnel
scale, $\sim \left( \lambda D / 2\pi \right)^{1/2}$, then the
fraction of the flux density in the scattered component 
$\sim \sigma_{\phi}^2$.   

The transition frequency between
weak and strong scattering may be evaluated (for Galactic sources) in terms of
the scattering measure, $\SM$, using an expression in
Cordes \& Lazio (2002),
\be
\nutrans = 225\, {\rm GHz}\, \SM^{6/17} \Deff^{5/17},
\label{eq:nutrans}
\ee
where $\Deff$ is the effective distance to the scattering medium.
For pulsars near the solar system with
$\log \SM \approx -4$ and $D\approx 0.1$ kpc, $\nutrans \approx 4.4$ GHz.
In the transition regime, scintillations behave similar to those
in the strong scattering regime.


\newpage
\begin{thebibliography}{}
\bibitem[Aller et al. 1985]{aller85} Aller,
H.~D., Aller, M.~F., Latimer, G.~E., \& Hodge, P.~E.\ 1985, \apjs, 59, 513
\bibitem[Amy et al. 1989]{amy89} Amy, S.\ W.,
Large, M.\ I., \& Vaughan, A.\ E.\ 1989, Proceedings of the Astronomical
Society of Australia, 8, 172
\bibitem[Aubier et al. 2000]{aubier2000} Aubier, A., Boudjada,
M.~Y., Moreau, P., Galopeau, P.~H.~M., Lecacheux, A., \& Rucker, H.~O.\
2000, \aap, 354, 1101
\bibitem[Backer et al.(1997)]{1997PASP..109...61B} Backer, D.~C., Dexter, 
M.~R., Zepka, A., Ng, D., Werthimer, D.~J., Ray, P.~S., \& Foster, R.~S.\ 
1997, \pasp, 109, 61
\bibitem[Berger et al. 2001]{berger01} Berger, E.~et al.\ 2001,
\nat, 410, 338
\bibitem[Bhat, Rao, \& Gupta 1999]{brg99}
        Bhat, N.~D.~R., Rao, A.~P., \& Gupta, Y.  1999, \apjs, 121, 483
\bibitem[Bogdanov et al. 2002] {bo2002} Bogdanov, S., Pruszy{\' n}ska, M.,
Lewandowski, W., \& Wolszczan, A.\ 2002, \apj, 581, 495
\bibitem[Camilo \& Nice 1995]{camilo1995} Camilo, F.~\& Nice,
D.~J.\ 1995, \apj, 445, 756
\bibitem[Cognard et al. 1996]{cognard96}
Cognard, I., Shrauner, J.\ A., Taylor, J.\ H., \& Thorsett, S.\ E.\ 1996,
\apjl, 457, L81
\bibitem[Cohen \& Brebner 1985]{cohen1985} Cohen, R.~J.~\& Brebner, G.~C.\ 1985, \mnras, 216, 51P
\bibitem[Colgate \& Noerdlinger 1971]{colgate71} Colgate, S.\
A.\ \& Noerdlinger, P.\ D.\ 1971, \apj, 165, 509
\bibitem[Cordes~1986]{c86} Cordes, J.~M.  1986, \apj, 311, 183
\bibitem[Cordes \& Lazio 1991]{cordes1991} Cordes, J.~M.~\& Lazio, T.~J.\ 1991, \apj, 376, 123
\bibitem[Cordes et al. 1997]{cordes1997} Cordes, J.~M.,
Lazio, T.~J.~W., \& Sagan, C.\ 1997, \apj, 487, 782
\bibitem[Cordes \& Chernoff 1998]{cc1998} Cordes, J.~M.~\& Chernoff, D.~F.\ 1998, \apj, 505, 315
\bibitem[Cordes \& Lazio 2002]{cl02}
Cordes, J. M. \& Lazio, T. J. W. L. 2002, astro-ph/0207156
\bibitem[Cortiglioni et al. 1981]{cort81} Cortiglioni, S.,
Mandolesi, N., Morigi, G., Ciapi, A., Inzani, P., \& Sironi, G.\ 1981,
\apss, 75, 153
\bibitem[Davies \& Large 1970]{davies70} Davies, J.\ G., \& Large, M.\ I., 1970, MNRAS, 149, 301
\bibitem[Dewey, Cordes, Wolszczan, \& Weisberg(1988)]{1988rwsi.conf..217D} 
Dewey, R.~J., Cordes, J.~M., Wolszczan, A., \& Weisberg, J.~M.\ 1988, AIP 
Conf.~Proc.~174: Radio Wave Scattering in the Interstellar Medium, 217 
\bibitem[Edwards et al. 1974]{edwards74} Edwards, P.\ J., Hurst, R.\ B., \& McQueen, M.\ P.\ C., 1974, \nat, 247, 444
\bibitem[Foster et al. 1991]{foster1991} Foster,
R.~S., Fairhead, L., \& Backer, D.~C.\ 1991, \apj, 378, 687
\bibitem[Fruchter et al. 1988]{fruch1988} Fruchter, A.~S.,
Taylor, J.~H., Backer, D.~C., Clifton, T.~R., \& Foster, R.~S.\ 1988, \nat,
331, 53
\bibitem[Goodman 1997]{goodman1997} Goodman, J.\ 1997, New
Astronomy, 2, 449
\bibitem[Gothoskar \& Gupta 2000]{go2000} Gothoskar, P.~\&
Gupta, Y.\ 2000, \apj, 531, 345
\bibitem[Hansen \& Lyutikov(2001)]{2001MNRAS.322..695H} Hansen, B.~M.~S.~\& 
Lyutikov, M.\ 2001, \mnras, 322, 695 
\bibitem[Hankins 1971]{Hankins1971} Hankins, T.~H.\ 1971, \apj,
169, 487
\bibitem[Hankins \& Rickett 1975]{hankins75} Hankins, T.\ H.\ \&
Rickett, B.\ J.\ 1975, Methods in Computational Physics.\ Volume 14 - Radio
astronomy, 14, 55
\bibitem[Hankins et al. 1981]{hankins81} Hankins, T.\ H.\ et
al.\ 1981, \apjl, 244, L61
\bibitem[Hankins 2003]{hankins03} Hankins, T.\ H.\,  et al. 
2003, \nat, 422, 141.
\bibitem[Harris, Zeissig, \& Lovelace(1970)]{1970A&A.....8...98H} Harris, 
D.~E., Zeissig, G.~A., \& Lovelace, R.~V.\ 1970, \aap, 8, 98 
\bibitem[Hesse \& Wielebinski 1974]{hesse74} Hesse, K.\ H.\ \&
Wielebinski, R.\ 1974, \aap, 31, 409
\bibitem[Hewish et al. 1968]{hewish68} Hewish, A., Bell-Burnell, J., Pilkington, J.\ D.\ H., Scott, P.\ F., \& Collins, R.\ A, 1968, {it Nature}, 217, 709
\bibitem[Horowitz \& Sagan(1993)]{1993ApJ...415..218H} Horowitz, P.~\& 
Sagan, C.\ 1993, \apj, 415, 218 
\bibitem[Hughes \& Retallack 1973]{hughes73} Hughes, V. A. \& Retallack,
D. S., 1973, {\it Nature}, 242, 105
\bibitem[Huguenin \& Moore 1974]{hug74} Huguenin, G.\ R.\ \&
Moore, E.\ L.\ 1974, \apjl, 187, L57
\bibitem[Inzani et al. 1982]{inzani82}
Inzani, P., Sironi, G., Mandolesi, N., \& Morigi, G.\ 1982, AIP Conf.\
Proc.\ 77: Gamma Ray Transients and Related Astrophysical Phenomena, 79
\bibitem[Jackson et al. 1989]{jackson1989} Jackson,
P.~D., Kundu, M.~R., \& White, S.~M.\ 1989, \aap, 210, 284
\bibitem[Johnston et al. 1998]{johnston1998}
Johnston, S., Nicastro, L., \& Koribalski, B.\ 1998, \mnras, 297, 108
\bibitem[Johnston \& Romani 2003]{jr03}
Johnston, S. \& Romani, R. W. 2003, \apj, submitted.
\bibitem[Kardashev et al. 1977]{kardashev77} Kardashev, N.\ S.\ et
al.\ 1977, Soviet Astronomy, 21, 1
\bibitem[Kedziora-Chudczer et al. 2001]{ked2001}
Kedziora-Chudczer, L.~L., Jauncey, D.~L., Wieringa, M.~H., Tzioumis, A.~K.,
\& Reynolds, J.~E.\ 2001, \mnras, 325, 1411
\bibitem[Lainela 1994]{lain1994} Lainela, M.\ 1994, \aap, 286,
408
\bibitem[Lambert \& Rickett(1999)]{LR1999} Lambert, H.~C.~\&
Rickett, B.~J.\ 1999, \apj, 517, 299.
\bibitem[Lecacheux et al. 1998]{lec1998} Lecacheux, A.,
Boudjada, M.~Y., Rucker, H.~O., Bougeret, J.~L., Manning, R., \& Kaiser,
M.~L.\ 1998, \aap, 329, 776
\bibitem[Linscott \& Erkes 1980]{linscott80} Linscott, I.\ R.\ \&
Erkes, J.\ W.\ 1980, \apjl, 236, L109
\bibitem[L{\" o}hmer et al. 2001]{lohmer2001} L{\" o}hmer, O.,
Kramer, M., Mitra, D., Lorimer, D.~R., \& Lyne, A.~G.\ 2001, \apjl, 562,
L157.
\bibitem[Mann et al. 1996]{mann1996} Mann, G., Klassen, A.,
Classen, H.-T., Aurass, H., Scholz, D., MacDowall, R.~J., \& Stone, R.~G.\
1996, \aaps, 119, 489
\bibitem[McCulloch et al. 1981]{mcculloch81}
McCulloch, P.\ M., Ellis, G.\ R.\ A., Gowland, G.\ A., \& Roberts, J.\ A.\
1981, \apjl, 245, L99
\bibitem[Narayan, R. 1992]{nar1992}
Narayan, R. 1992, Phil. Trans. R. Soc. Lond. A, 341, 151.
\bibitem[Nicastro et al. 2001]{nic2001} Nicastro, L., Nigro,
F., D'Amico, N., Lumiella, V., \& Johnston, S.\ 2001, \aap, 368, 1055
\bibitem[Nice et al. 1995]{nice95} Nice, D.~J.,
Fruchter, A.~S., \& Taylor, J.~H.\ 1995, \apj, 449, 156
\bibitem[Nice 1999]{nice99} Nice, D.\ J.\ 1999, \apj, 513,
927
\bibitem[O'Sullivan et al. 1978]{osullivan78} O'Sullivan,
J.\ D., Ekers, R.\ D., \& Shaver, P.\ A.\ 1978, \nat, 276, 590
\bibitem[Phillips \& Clegg 1992]{phil1992} Phillips, J.~A.~\&
Clegg, A.~W.\ 1992, \nat, 360, 137
\bibitem[Phinney \& Taylor 1979]{phinney79} Phinney, S.\ \&
Taylor, J.\ H.\ 1979, \nat, 277, 117
\bibitem[Poquerusse et al. 1988]{poq1988} Poquerusse, M.,
Steinberg, J.~L., Caroubalos, C., Dulk, G.~A., \& MacQueen, R.~M.\ 1988,
\aap, 192, 323
\bibitem[Rees 1977]{rees77} Rees, M.\ J.\ 1977, \nat, 266,
333
\bibitem[Rickett 1990]{ric90} Rickett, B.~J., 1990, \araa, 28, 561
\bibitem[Romani \& Johnston 2001]{romani2001} Romani, R.~W.~\& Johnston, S.\ 2001, \apjl, 557, L93.
\bibitem[Staelin \& Reifenstein 1968]{sta68} Staelin, D. H. \& Reifenstein, E.\ C., 1968, {\it Science}, 162, 1481
\bibitem[Taylor et al. 1972]{taylor72} Taylor,
J.\ H., Huguenin, G.\ R., \& Hirsch, R.\ M.\ 1972, \apjl, 172, L17
\bibitem[Taylor et al. 1981]{taylor81} Taylor,
J.\ H., Backus, P.\ R., \& Damashek, M.\ 1981, \apjl, 244, L65
\bibitem[Taylor et. al 1993]{ppcat} Taylor, J. H., Manchester, R. N., \& Lyne, A. G., 1993, ApJS,
88, 529
\bibitem[Turtle et al. 1987]{turtle1987} Turtle, A.~J.,
Campbell-Wilson, D., Bunton, J.~D., Jauncey, D.~L., \& Kesteven, M.~J.\
1987, \nat, 327, 38
\bibitem[Weber 1969]{weber69} Weber, J. 1969, {\it Phys. Rev. Letters}, 22, 1320
\bibitem[Williamson(1972)]{will72} Williamson, I.~P.\ 1972,
\mnras, 157, 55.
\bibitem[Yudaeva 1986]{yud1986} Yudaeva, N.~A.\ 1986, Soviet
Astronomy Letters, 12, 150
\end{thebibliography}
\end{document}